\documentclass[aps,prb,reprint,superscriptaddress]{revtex4-2}

\usepackage{amsmath,amssymb,bm,amsthm}
\DeclareMathOperator{\Supp}{Supp}
\usepackage{graphicx}
\usepackage{booktabs}
\usepackage{placeins}
\newcommand{\RaggedCell}[1]{{\raggedright #1\par}}
\usepackage{xcolor}
\usepackage{tikz}
\usetikzlibrary{positioning, arrows.meta, shapes.geometric}

\usepackage[utf8]{inputenc}
\usepackage{CJKutf8}

\usepackage[
  colorlinks=true,
  allcolors=blue,
  pdftitle={Parent-Group Generation of Linear Magnetoelectric Response: Two-Tier Screening and Completeness Classification},
  pdfauthor={Yachao Liu, Vei Wang, and Haibo Niu}
]{hyperref}

\newtheorem{theorem}{Theorem}
\newtheorem{lemma}{Lemma}
\newtheorem{corollary}{Corollary}

\definecolor{mered}{HTML}{D32F2F}
\definecolor{meorange}{HTML}{E65100}
\definecolor{meblue}{HTML}{1976D2}

\newcommand{\cR}[1]{\textcolor{mered}{\textbf{\boldmath #1}}}    
\newcommand{\cO}[1]{\textcolor{meorange}{\textbf{\boldmath #1}}} 
\newcommand{\cB}[1]{\textcolor{meblue}{\textbf{\boldmath #1}}}   

\begin{document}

\begin{CJK*}{UTF8}{gbsn}

\title{Parent-Group Generation of Linear Magnetoelectric Response: Two-Tier Screening and Completeness Classification}

\author{Yachao Liu (刘雅超)}
\email{liuyachao@xaut.edu.cn}
\affiliation{Department of Applied Physics, School of Optoelectronic Science and Intelligent Instrumentation, Xi'an University of Technology, Xi'an 710048, China}
\affiliation{Interdisciplinary Research Center for Semiconductor and Artificial Intelligence, Xi'an University of Technology, Xi'an 710048, China}

\author{Vei Wang (王伟)}
\affiliation{Department of Electronic Engineering, Xi'an University of Technology, Xi'an 710048, China}
\affiliation{Interdisciplinary Research Center for Semiconductor and Artificial Intelligence, Xi'an University of Technology, Xi'an 710048, China}

\author{Haibo Niu (牛海波)}
\email{n\_haibo@163.com}
\affiliation{Department of Applied Physics, Xi'an Jiaotong University City College, Xi'an 710018, China}

\date{\today}

\begin{abstract}
Given the magnetic atomic sites of a crystal and a $\mathbf{k}=0$ magnetic ordering pattern, which net bulk linear magnetoelectric (ME) tensor components are allowed, how many independent coupling coefficients govern them, and how does the response rotate with the magnetic order? Here we formulate an upstream two-tier theory for the homogeneous, static linear ME response to first order in a single irreducible magnetic mode across all 32 crystallographic point groups. Rather than constraining property tensors only downstream from a fixed magnetic point group (MPG), the unperturbed parent point group $G$ specifies the universal nine-dimensional electromagnetic coupling space while the magnetic Wyckoff sublattice $\mathcal{S}_{\mathrm{mag}}$ determines which spin arrangements can occur. Two-tier screening identifies ME-active modes, and parent-group invariant construction yields explicit leading-order tensor templates $\alpha_{ij}^{(1)}(\mathbf{L})$ resolved into monopolar, toroidal, and quadrupolar sectors. Evaluating these templates at a condensed order-parameter direction embeds the response into the residual MPG tensor space---either filling it or forming a strict subspace (such as 3D within 5D) enforcing additional leading-order component relations---while linking all orientations into a single parameter-sparse tensor family. Applying this workflow to $\mathrm{Cr_2O_3}$, $\alpha\text{-}\mathrm{Fe_2O_3}$, $\mathrm{BiFeO_3}$, and $\mathrm{Cu_2OSeO_3}$ contrasts the diagonal and forbidden corundum responses, identifies a purely antisymmetric fingerprint for a candidate $A_{2u}$ ordering on the same $4c$ sublattice, and demonstrates few-parameter angular locking across in-plane and three-dimensional magnetic rotations. Globally, all 32 parent groups partition into 14 ME-complete groups forming a subgroup down-set and 18 mode-selective groups, establishing the equivalence of leading-order and residual-MPG whole-mode symmetry prohibition and explaining why anti-inversion symmetry alone is insufficient.
\end{abstract}

\maketitle
\end{CJK*}
\raggedbottom

\section{Introduction}

The linear magnetoelectric (ME) effect couples electric and magnetic responses in quantum materials, offering both a route toward low-power spintronic and multiferroic control and a sensitive probe of magnetic symmetry breaking \cite{Fiebig2005,Eerenstein2006,SpaldinRamesh2019,Mostovoy2024,Spaldin2026}. Its static free-energy contribution is written as $F_{\mathrm{me}} = -\sum_{ij}\alpha_{ij}(\mathbf{L})E_i H_j$, where $\mathbf{E}$ is the polar electric field, $\mathbf{H}$ is the axial magnetic field, and $\alpha_{ij}(\mathbf{L})$ is the macroscopic linear ME tensor controlled by a magnetic order parameter $\mathbf{L}$. A classic physical puzzle immediately illustrates why crystal symmetry must be combined with atomic-site spin arrangements: isostructural corundum antiferromagnets $\mathrm{Cr_2O_3}$ and $\alpha\text{-}\mathrm{Fe_2O_3}$ share the exact same parent space group ($R\bar{3}c$) and $4c$ magnetic Wyckoff sublattice, yet $\mathrm{Cr_2O_3}$ exhibits a classic, measurable bulk linear ME effect \cite{Dzyaloshinskii1960,Astrov1961}, whereas ideal inversion-preserving bulk phases of $\alpha\text{-}\mathrm{Fe_2O_3}$ have zero net bulk static linear ME response---even in the weakly ferromagnetic phase above the Morin transition---while still permitting mutually compensating site-local responses \cite{Morin1950,lebRun2020,Verbeek2023}. This contrast motivates a central practical question for theory, computation, and experiment: given the magnetic atomic positions of a crystal and a homogeneous ($\mathbf{k}=0$) magnetic ordering mode, which linear ME components are symmetry-allowed, how many independent coupling coefficients exist, and how do the observable polarization direction and tensor components evolve as the magnetic order rotates?

Because $\alpha_{ij}$ is odd under both spatial inversion ($\mathcal{I}$) and time reversal ($\mathcal{T}$), its established $1+3+5$ decomposition separates a pseudoscalar monopolar trace $\theta$, an antisymmetric toroidal response vector $\mathbf{T}_\alpha$, and a symmetric-traceless quadrupolar tensor $\mathbf{Q}_S$ \cite{SpaldinFiebigMostovoy2008,Hehl2009,WatanabeYanase2018}. These macroscopic response coefficients share the transformation properties of microscopic magnetoelectric multipoles without being numerically identical to them \cite{SpaldinFiebigMostovoy2008}, and the isotropic trace $\theta$ is distinct from the topological axion angle \cite{Essin2009,Qi2008}. While linear magnetoelectricity connects broadly to non-reciprocal optical and transport phenomena \cite{TokuraNagaosa2018,NagaosaYanase2024}, magnetic multipolar order \cite{Hayami2018,Yatsushiro2021,Hayami2024}, and antiferromagnetic spintronics \cite{Jungwirth2016,Rimmler2025}, those dynamical or nonlinear phenomena carry process-dependent symmetry conditions; here we focus specifically on the net bulk, homogeneous, static linear ME tensor $\alpha_{ij}$ to first order in a single irreducible $\mathbf{k}=0$ magnetic mode.

Established symmetry frameworks address different parts of this problem. Representation analysis (RA) classifies candidate spin configurations and invariant couplings under the paramagnetic parent group \cite{Bertaut1968,Dresselhaus2008}, whereas magnetic-point-group (MPG) analysis constrains the property tensors of a specified ordered phase through its invariant subspace $\operatorname{Fix}_M(V_{\mathrm{ME}})$ \cite{LandauEDC,Rivera1994,Schmid1994,Litvin2014}, with crystallographic tools linking the two descriptions via order-parameter directions (OPDs), isotropy subgroups, and tensor tabulations (e.g., \textsc{Get\_mirreps} and \textsc{MTensor}) \cite{RodriguezCarvajal2024,Tasci2025}. Systematic tensorial-covariant methods construct tensor bases resolved by parent-group irreps, resolve multiplicities, aligned partners, and parity relations, and convert them into Cartesian property tensors along symmetry descents \cite{Kopsky2006,Kopsky2006b}; virtual-cluster multipole schemes and symmetry-adapted microscopic modeling construct complete $\mathbf{k}=0$ magnetic bases, incorporate Wyckoff-site realizability, and address field couplings and response selection \cite{Suzuki2019,Kusunose2023}; and multipolar classifications across all 32 crystallographic and 122 magnetic point groups connect parent irreps to linear response tensors, including the scalar and quadrupolar channels of $\mathrm{Cr_2O_3}$ and parity-odd ME-inactive phases \cite{Hayami2018,WatanabeYanase2018,Yatsushiro2021,Hayami2024}. Concurrently, parent-symmetry Landau expansions have enumerated multiferroic couplings in $Pm\bar{3}m$ perovskites \cite{SennBristowe2018}, derived sublattice-mode-resolved ME tensors in perovskite and Ruddlesden--Popper families \cite{Gareeva2022} (reproducing the $\mathrm{BiFeO_3}$ tensor attributed to Ref.~\cite{Popov1994}), classified inhomogeneous spin-gradient couplings across all 32 point groups \cite{Tu2026}, and established automated parent-space-group expansions of response tensors in continuous magnetic order parameters, including the signs associated with antiferromagnetic sublattice exchange \cite{Xiao2026}.

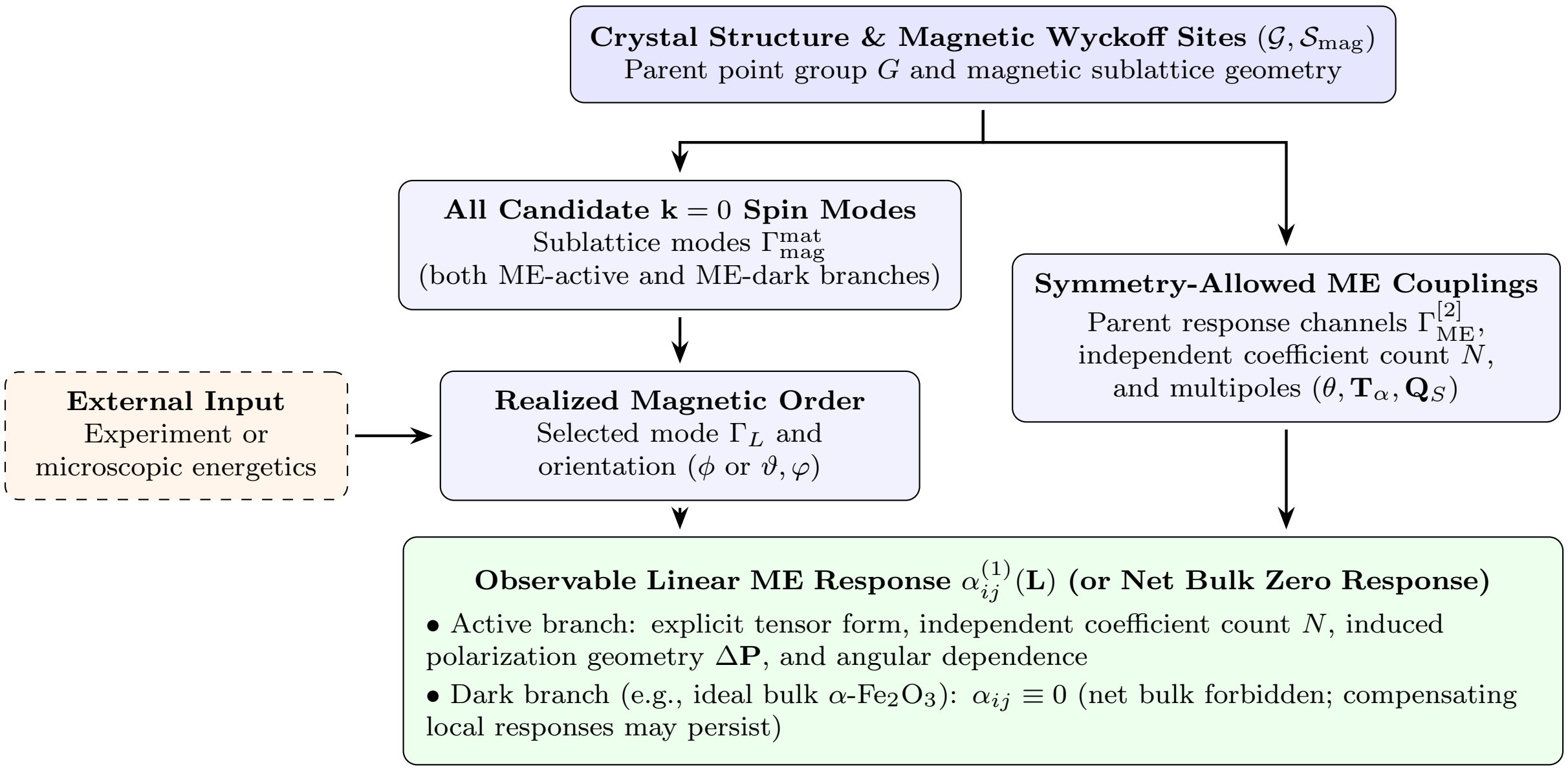
\begin{figure*}[!t]
\centering
\begin{tikzpicture}[
  >=Stealth,
  box/.style={draw, rounded corners=4pt, fill=blue!5, inner sep=5.5pt, font=\footnotesize, align=center},
  ext/.style={draw, dashed, rounded corners=4pt, fill=orange!8, inner sep=5.5pt, font=\footnotesize, align=center},
  outbox/.style={draw, rounded corners=4pt, fill=green!7, inner sep=6.5pt, font=\footnotesize, align=left},
  arrow/.style={->, thick, shorten >=2pt, shorten <=2pt}
]
\node[box, fill=blue!10, minimum width=8.2cm] (crystal) at (1.8, 3.90) {%
\textbf{Crystal Structure \& Magnetic Wyckoff Sites} ($\mathcal{G}, \mathcal{S}_{\mathrm{mag}}$)\\
Parent point group $G$ and magnetic sublattice geometry};

\node[box, minimum width=4.9cm] (modes) at (-1.3, 1.95) {%
\textbf{All Candidate $\mathbf{k}=0$ Spin Modes}\\
Sublattice modes $\Gamma_{\mathrm{mag}}^{\mathrm{mat}}$\\
(both ME-active and ME-dark branches)};

\node[ext, minimum width=3.5cm] (exp) at (-6.45, 0.0) {%
\textbf{External Input}\\
Experiment or\\
microscopic energetics};

\node[box, minimum width=4.9cm] (realized) at (-1.3, 0.0) {%
\textbf{Realized Magnetic Order}\\
Selected mode $\Gamma_L$ and\\
orientation ($\phi$ or $\vartheta,\varphi$)};

\node[box, minimum width=5.6cm] (couplings) at (4.9, 1.00) {%
\textbf{Symmetry-Allowed ME Couplings}\\
Parent response channels $\Gamma_{\mathrm{ME}}^{[2]}$,\\
independent coefficient count $N$,\\
and multipoles ($\theta, \mathbf{T}_\alpha, \mathbf{Q}_S$)};

\node[outbox, text width=11.4cm] (output) at (1.8, -2.20) {%
{\centering\textbf{Observable Linear ME Response $\alpha_{ij}^{(1)}(\mathbf{L})$ (or Net Bulk Zero Response)}\par\vspace{3pt}}%
{\raggedright
$\bullet$ Active branch: explicit tensor form, independent coefficient count $N$, induced polarization geometry $\Delta\mathbf{P}$, and angular dependence\par\vspace{1.5pt}
$\bullet$ Dark branch (e.g., ideal bulk $\alpha\text{-}\mathrm{Fe_2O_3}$): $\alpha_{ij} \equiv 0$ (net bulk forbidden; compensating local responses may persist)\par}};

\draw[arrow] (crystal.south) -- (1.8, 3.00) -| (modes.north);
\draw[arrow] (crystal.south) -- (1.8, 3.00) -| (couplings.north);

\draw[arrow] (modes.south) -- (realized.north);
\draw[arrow] (exp.east) -- (realized.west);
\draw[arrow] (realized.south) -- (output.north -| realized.south);
\draw[arrow] (couplings.south) -- (output.north -| couplings.south);
\end{tikzpicture}
\caption{Physical workflow of parent-group linear magnetoelectric response generation. From the crystal structure and magnetic Wyckoff sites, two parallel inputs converge: (left) all candidate $\mathbf{k}=0$ magnetic modes on the sublattice, from which experiment or microscopic energetics selects the realized magnetic order and orientation (naturally retaining ME-dark phases such as $\alpha\text{-}\mathrm{Fe_2O_3}$); and (right) the symmetry-allowed parent-group ME coupling channels and independent coefficient counts. Combining them yields either the explicit leading-order tensor $\alpha_{ij}^{(1)}(\mathbf{L})$, its polarization geometry and angular relations, or a symmetry-enforced zero net bulk response. A technical representation-theoretic summary of the projection and intertwiner machinery is provided in Fig.~\ref{fig:tech_summary} of Appendix~\ref{app:category}.}
\label{fig:framework}
\end{figure*}

Building on these foundations and on the site-resolved polar--axial mapping of Part~I \cite{Liu_mapping} (with third-rank piezomagnetism treated in Part~III \cite{Liu_piezo}), we establish a unified upstream two-tier framework for homogeneous $\mathbf{k}=0$ linear ME response across all 32 crystallographic point groups that systematically complements downstream MPG constraints. Whereas a residual MPG table specifies only the static symmetry envelope of an already fixed ordered phase, retaining the parent-group irrep identity and sublattice kinematics upstream generates the explicit leading-order tensor family $\alpha_{ij}^{(1)}(\mathbf{L})$, links one-dimensional real sign irreps isomorphically to their induced MPG tensor spaces, and determines whether a multidimensional mode fills the residual MPG space or forms a strict leading-order subspace (such as 3D within 5D) enforcing sum rules and angular locking under magnetic rotation---while organizing the global 14/18 parent-completeness hierarchy that is not explicit in individual magnetic-group tables. As illustrated in Fig.~\ref{fig:framework}, the crystal structure and magnetic Wyckoff sublattice define all candidate $\mathbf{k}=0$ spin modes (from which experiment or microscopic energetics selects the realized magnetic order and orientation), while the parent point group determines which modes can couple linearly to $\mathbf{E}$ and $\mathbf{H}$ and how many independent coupling coefficients they support, yielding either $\alpha_{ij}^{(1)}(\mathbf{L})$ and its directional evolution or a symmetry-enforced net bulk ME-dark state.

The remainder of this paper is organized as follows. Section~II develops the constitutive channel construction, the polar--axial pseudoscalar simplification, and the multipolar decomposition. Section~III applies the standardized workflow to four representative systems: contrasting $\mathrm{Cr_2O_3}$ and $\alpha\text{-}\mathrm{Fe_2O_3}$ to isolate the role of inversion parity and weak ferromagnetism, analyzing induced-polarization geometry and in-plane angular locking in $\mathrm{BiFeO_3}$ \cite{CatalanScott2009,Wang2003}, and examining three-dimensional orientation dependence, probe-field selectivity, and magnetic pinning in chiral $\mathrm{Cu_2OSeO_3}$ \cite{Bos2008,Seki2012}. Section~IV presents the global 14/18 parent-completeness classification as a practical screening tool, together with the parent-representation origin and all-odd-order persistence of single-mode dark channels. Throughout, symmetry determines allowed tensor templates and directional relations, while numerical coupling amplitudes require microscopic electronic, spin-orbit, and lattice-mediated calculations \cite{Xu2024,Iniguez2008,WojdelIniguez2009}.

\section{Parent-Group Formalism and Response Generation}

\subsection{Paramagnetic Gray Parent Group and Response Space}

In the paramagnetic parent phase, macroscopic point symmetry is described by the direct-product gray parent group $\mathcal{G}_{\mathrm{P}} = G \times \Theta_{\mathcal{T}} = G \cup \mathcal{T}G$, where $G = \mathcal{G}/\mathcal{T}_{\mathrm{lat}}$ is the unitary spatial crystallographic point group of the parent space group $\mathcal{G}$ and $\Theta_{\mathcal{T}} = \{E, \mathcal{T}\}$ is the time-reversal group \cite{Dresselhaus2008,RodriguezCarvajal2024,Liu_mapping}. The \emph{spatiotemporal separation principle} dictates that the spatial point group $G$ governs the geometric configuration of physical modes, while $\Theta_{\mathcal{T}}$ independently assigns a time-reversal parity grading $\tau_{\mathcal{T}} \in \{+1, -1\}$.

The fundamental geometric duality between polar and axial spatial quantities across all 32 crystallographic point groups is governed by the canonical one-dimensional pseudoscalar representation $\Gamma_{\mathrm{ps}}$, whose character is the spatial transformation determinant \cite{Liu_mapping}:
\begin{equation}
\chi_{\mathrm{ps}}(g) = \det(R_g) \in \{+1, -1\}, \quad \forall g \in G.
\label{eq:pseudoscalar_def}
\end{equation}
Lattice displacements transform as polar, time-even vectors ($\Gamma_{\mathrm{pol}}, \tau_{\mathcal{T}}=+1$), whereas localized magnetic moments and external magnetic fields transform as axial, time-odd vectors ($\Gamma_{\mathrm{axial}} = \Gamma_{\mathrm{pol}} \otimes \Gamma_{\mathrm{ps}}, \tau_{\mathcal{T}}=-1$) \cite{Liu_mapping}. Under external fields $\mathbf{E} \sim (\Gamma_{\mathrm{pol}}, \tau_{\mathcal{T}}=+1)$ and $\mathbf{H} \sim (\Gamma_{\mathrm{axial}}, \tau_{\mathcal{T}}=-1)$, the linear magnetoelectric response tensor space $V_{\mathrm{ME}}$ carries time-reversal-odd parity $\tau_{\mathcal{T}}(V_{\mathrm{ME}}) = (+1)(-1) = -1$ and a nine-dimensional spatial representation $\Gamma_{\mathrm{ME}}^{[2]}$ of rank-two polar--axial tensors:
\begin{equation}
\Gamma_{\mathrm{ME}}^{[2]} \equiv \Gamma_{\mathrm{pol}} \otimes \Gamma_{\mathrm{axial}} = \Gamma_{\mathrm{pol}}^{\otimes 2} \otimes \Gamma_{\mathrm{ps}} = \bigoplus_{\rho \in \mathrm{Irr}(G)} n_\rho^{\mathrm{ME}} \, \rho,
\label{eq:MEspace}
\end{equation}
where $n_\rho^{\mathrm{ME}} = \langle \rho, \Gamma_{\mathrm{ME}}^{[2]} \rangle$ is the multiplicity of the active irrep $\rho$ in the response space, defining the parent-group ME support $\mathcal{S}_{\mathrm{ME}}(G) \equiv \Supp(\Gamma_{\mathrm{ME}}^{[2]}) = \{ \rho \in \mathrm{Irr}(G) \mid n_\rho^{\mathrm{ME}} > 0 \}$. Because $\Gamma_{\mathrm{pol}}$ and $\Gamma_{\mathrm{axial}}$ are real orthogonal representations, $\Gamma_{\mathrm{ME}}^{[2]}$ is a real representation ($V_{\mathrm{ME}}^* \cong V_{\mathrm{ME}}$) with a real-valued character $\chi_{\mathrm{ME}}(g) \in \mathbb{R}$, ensuring $n_{\rho^*}^{\mathrm{ME}} = n_\rho^{\mathrm{ME}}$ for every $\rho \in \mathrm{Irr}(G)$.

We distinguish this universal parent response space $V_{\mathrm{ME}} \equiv \Gamma_{\mathrm{ME}}^{[2]}$ below from the material-specific sublattice magnetic representation $\Gamma_{\mathrm{mag}}^{\mathrm{mat}}$, the downstream invariant tensor subspace $\operatorname{Fix}_M(V_{\mathrm{ME}})$ permitted by a symmetry-broken magnetic point group, and the material-specific realization of the corresponding coupling amplitudes.

\subsection{Pseudoscalar Cancellation, Two-Tier Screening, and Constitutive Generation}

The static free-energy coupling between a condensed homogeneous magnetic order parameter $\mathbf{L} \sim (\Gamma_L, \tau_{\mathcal{T}}(L)=-1)$ and external electric and magnetic fields is written as a function of $\mathbf{L}$:
\begin{equation}
F_{\mathrm{me}} = -\sum_{ij} \alpha_{ij}(\mathbf{L}) E_i H_j.
\label{eq:parent_free_energy}
\end{equation}
Retaining the general function $\alpha_{ij}(\mathbf{L})$ distinguishes linearity in the probe fields $\mathbf{E}, \mathbf{H}$ from the expansion order in $\mathbf{L}$ \cite{Mostovoy2024} and leaves room for higher-order extensions. In this work, we construct the direct leading-order contribution $\alpha_{ij}^{(1)}(\mathbf{L})$ linear in $\mathbf{L}$, with the magnetic mode held fixed when differentiating with respect to the probe fields. At first order in $\mathbf{L}$, an internal trilinear Landau coupling $\gamma_{ij\mu}L_\mu p_i m_j$ describes the same symmetry-allowed channels when $\mathbf{p}$ and $\mathbf{m}$ are independent polarization and magnetization response fluctuations around equilibrium with regular, nonsingular harmonic susceptibilities ($\chi^e, \chi^m$); eliminating these fluctuations yields $\alpha_{ij}^{(1)}(\mathbf{L}) = -\chi_{ii'}^e \gamma_{i'j'\mu} L_\mu \chi_{j'j}^m$ in a consistent conjugate-field convention. (If $\mathbf{L}$ itself contains a ferromagnetic component, $\mathbf{m}$ denotes a response fluctuation about that condensed background rather than a second copy of the total magnetization.)

Invariance of the trilinear pairing in Eq.~\eqref{eq:parent_free_energy} under the full gray parent group $\mathcal{G}_{\mathrm{P}} = G \times \Theta_{\mathcal{T}}$ requires the dual screening conditions \cite{Liu_mapping}:
\begin{equation}
\begin{cases}
\Gamma_L^* \otimes \Gamma_{\mathrm{pol}} \otimes \Gamma_{\mathrm{axial}} \supset \Gamma_1, & (G\text{-invariance}), \\[3pt]
\tau_{\mathcal{T}}(L)\,\tau_{\mathcal{T}}(E)\,\tau_{\mathcal{T}}(H) = +1, & (\Theta_{\mathcal{T}}\text{-invariance}),
\end{cases}
\label{eq:dual_criterion}
\end{equation}
where $\Gamma_L^*$ denotes the contragredient (dual) representation of $\Gamma_L$, which coincides with $\Gamma_L$ ($\Gamma_L^* \cong \Gamma_L$) for all real orthogonal representations. The time-reversal condition $(-1)(+1)(-1) = +1$ is identically satisfied for any time-odd magnetic order. In the spatial sector, expressing $\Gamma_L$ via the pseudoscalar correspondence with a polar representation $\Gamma$ ($\Gamma_L = \Gamma \otimes \Gamma_{\mathrm{ps}}$ \cite{Liu_mapping}) reveals that the pseudoscalar factor of the magnetic order is exactly compensated by the axial character of the magnetic field ($\Gamma_H = \Gamma_{\mathrm{pol}} \otimes \Gamma_{\mathrm{ps}}$). Because $\Gamma_{\mathrm{ps}}^* \cong \Gamma_{\mathrm{ps}}$ and $\Gamma_{\mathrm{ps}}^{\otimes 2} \equiv \Gamma_1$, pairwise cancellation yields:
\begin{equation}
\begin{aligned}
(\Gamma^* \otimes \Gamma_{\mathrm{ps}}) \otimes \Gamma_{\mathrm{pol}} \otimes (\Gamma_{\mathrm{ps}} \otimes \Gamma_{\mathrm{pol}}) &= \Gamma^* \otimes \Gamma_{\mathrm{pol}}^{\otimes 2} \otimes \Gamma_{\mathrm{ps}}^{\otimes 2} \\
&\equiv \Gamma^* \otimes \Gamma_{\mathrm{pol}}^{\otimes 2}.
\end{aligned}
\end{equation}
Consequently, parent-group spatial response screening reduces directly to an ordinary polar tensor product:
\begin{equation}
\boxed{
\Gamma^* \otimes \Gamma_{\mathrm{pol}}^{\otimes 2} \supset \Gamma_1
},
\label{eq:polar_reduction}
\end{equation}
reducing to $\Gamma \otimes \Gamma_{\mathrm{pol}}^{\otimes 2} \supset \Gamma_1$ whenever $\Gamma$ is self-dual.

\textit{Untwisted Channel-Count Identity.---}
Because $\Gamma_{\mathrm{pol}}^{\otimes 2}$ is a real orthogonal representation, this compensation establishes an exact identity for the constitutive channel multiplicity:
\begin{equation}
\boxed{
N(\Gamma_L) = \langle \Gamma_L, \Gamma_{\mathrm{ME}}^{[2]} \rangle = \langle \Gamma, \Gamma_{\mathrm{pol}}^{\otimes 2} \rangle
}.
\label{eq:untwisted_channel_count}
\end{equation}
The space of symmetry-allowed linear constitutive mappings forms the equivariant intertwiner space $\mathcal{M}_{\mathrm{ME}}(\Gamma_L) \equiv \operatorname{Hom}_G(\Gamma_L, \Gamma_{\mathrm{ME}}^{[2]}) \cong (V_{\mathrm{ME}} \otimes \Gamma_L^*)^G$, with dimension:
\begin{equation}
N = \dim_{\mathbb{C}} \mathcal{M}_{\mathrm{ME}}(\Gamma_L) = \langle \Gamma_L, \Gamma_{\mathrm{ME}}^{[2]} \rangle = n_{\Gamma_L}^{\mathrm{ME}}.
\label{eq:channel_number}
\end{equation}
For any real-type irreducible representation $\Gamma_L$ (including all condensed order-parameter sectors explicitly analyzed in our four benchmark materials below), $N$ in Eq.~\eqref{eq:channel_number} directly gives the number of linearly independent real Cartesian tensor templates; for a complex-type two-dimensional real doublet $E = \sigma \oplus \sigma^*$ formed by conjugate 1D irreps $\sigma, \sigma^* \in \mathrm{Irr}_{\mathbb{C}}(G)$, each doublet copy supports two real couplings and the total number of independent real parameters is $N(E) = \dim_{\mathbb{R}}\operatorname{Hom}_G^{\mathbb{R}}(E, V_{\mathrm{ME}}^{\mathbb{R}}) = n_\sigma^{\mathrm{ME}} + n_{\sigma^*}^{\mathrm{ME}} = 2n_\sigma^{\mathrm{ME}}$ (see \hyperref[app:proof]{Appendix~A}).

\textit{Material-Specific Two-Tier Screening.---}
While the parent-group support $\mathcal{S}_{\mathrm{ME}}(G) = \Supp(\Gamma_{\mathrm{ME}}^{[2]})$ establishes which irreducible representations of $G$ are universally permitted to couple linearly to $\mathbf{E}$ and $\mathbf{H}$ (tabulated across all 32 point groups in \hyperref[app:tables]{Appendix~B}), a specific crystal with parent space group $\mathcal{G}$ and a $G$-invariant magnetic Wyckoff sublattice $\mathcal{S}_{\mathrm{mag}}$ hosts a definite set of kinematically available $\mathbf{k}=0$ magnetic modes. As established in Part~I \cite{Liu_mapping}, let $Q_{\mathrm{mag}}: V_{\mathrm{ph}}^{\mathrm{full}} \to V_{\mathrm{ph}}^{\mathrm{mag}}$ denote the projection from the full $\Gamma$-point polar displacement space onto the magnetic sublattice $\mathcal{S}_{\mathrm{mag}}$. Because $\mathcal{S}_{\mathrm{mag}}$ is closed under $G$, $Q_{\mathrm{mag}}$ is $G$-equivariant ($Q_{\mathrm{mag}} D_{\mathrm{ph}}^{\mathrm{full}}(g) = D_{\mathrm{ph}}^{\mathrm{mag}}(g) Q_{\mathrm{mag}}, \, \forall g \in G$). Applying the pseudoscalar twist to the restricted polar representation $\Gamma_{\mathrm{ph}}\vert_{\mathcal{S}_{\mathrm{mag}}}$ reduced consistently to the point group $G = \mathcal{G}/\mathcal{T}_{\mathrm{lat}}$ yields the material-specific magnetic-sublattice representation:
\begin{equation}
\Gamma_{\mathrm{mag}}^{\mathrm{mat}} = \left( \Gamma_{\mathrm{ph}}\big\vert_{\mathcal{S}_{\mathrm{mag}}} \right) \otimes \Gamma_{\mathrm{ps}} = \bigoplus_{\rho \in \mathrm{Irr}(G)} m_\rho^{\mathrm{mat}} \, \rho,
\label{eq:sublattice_mag_rep}
\end{equation}
with sublattice kinematic support $\mathcal{S}_{\mathrm{mag}}^{\mathrm{mat}} \equiv \Supp(\Gamma_{\mathrm{mag}}^{\mathrm{mat}}) = \{ \rho \in \mathrm{Irr}(G) \mid m_\rho^{\mathrm{mat}} > 0 \}$. A magnetic order-parameter sector $\Gamma_L$ is both kinematically realizable on $\mathcal{S}_{\mathrm{mag}}$ and symmetry-allowed to generate a leading-order linear ME response if and only if it belongs to the two-tier support intersection (an intersection of irrep-type sets under the common parent point group $G$; see \hyperref[app:category]{Appendix~C} for the corresponding linear-mapping formulation):
\begin{equation}
\boxed{
\mathcal{S}_{\mathrm{act}}^{\mathrm{mat}} \equiv \Supp\left(\Gamma_{\mathrm{mag}}^{\mathrm{mat}}\right) \cap \Supp\left(\Gamma_{\mathrm{ME}}^{[2]}\right)
}.
\label{eq:two_tier_intersection}
\end{equation}
Combining this two-tier support screening with invariant construction defines the upstream constitutive chain:
\begin{equation}
\begin{aligned}
(\mathcal{G}, \mathcal{S}_{\mathrm{mag}}) &\longrightarrow \mathcal{S}_{\mathrm{act}}^{\mathrm{mat}} = \Supp\left(\Gamma_{\mathrm{mag}}^{\mathrm{mat}}\right) \cap \Supp\left(\Gamma_{\mathrm{ME}}^{[2]}\right) \\
&\xrightarrow{\Gamma_L \in \mathcal{S}_{\mathrm{act}}^{\mathrm{mat}}} \operatorname{Hom}_G\left(\Gamma_L, \Gamma_{\mathrm{ME}}^{[2]}\right) \longrightarrow \alpha_{ij}^{(1)}(\mathbf{L}).
\end{aligned}
\label{eq:generative_arrow}
\end{equation}
Relative to $\mathrm{Irr}(G)$ and the specified magnetic sublattice $\mathcal{S}_{\mathrm{mag}}$, candidate irreps partition into three nontrivial classes (with Class~D irreps absent from both supports excluded from the candidate space):
(i)~\emph{Class~A (Material-available and ME-active):} $\Gamma_L \in \mathcal{S}_{\mathrm{mag}}^{\mathrm{mat}} \cap \mathcal{S}_{\mathrm{ME}}(G)$, where both $m_{\Gamma_L}^{\mathrm{mat}} > 0$ and $n_{\Gamma_L}^{\mathrm{ME}} > 0$;
(ii)~\emph{Class~B (Material-available but linear-ME dark):} $\Gamma_L \in \mathcal{S}_{\mathrm{mag}}^{\mathrm{mat}} \setminus \mathcal{S}_{\mathrm{ME}}(G)$, where the magnetic sublattice hosts the mode ($m_{\Gamma_L}^{\mathrm{mat}} > 0$) but parent symmetry forbids an independent linear ME coupling at leading order ($n_{\Gamma_L}^{\mathrm{ME}} = 0$); and
(iii)~\emph{Class~C (Parent-allowed but sublattice-unavailable):} $\Gamma_L \in \mathcal{S}_{\mathrm{ME}}(G) \setminus \mathcal{S}_{\mathrm{mag}}^{\mathrm{mat}}$, where the parent point group permits linear ME coupling ($n_{\Gamma_L}^{\mathrm{ME}} > 0$) but the specified $\mathbf{k}=0$ magnetic Wyckoff sublattice lacks the kinematic degrees of freedom to host $\Gamma_L$ ($m_{\Gamma_L}^{\mathrm{mat}} = 0$). To illustrate Class~C in a single-orbit model, consider the ME-complete polar group $C_{3v}$ ($\mathcal{S}_{\mathrm{ME}}(C_{3v}) = \{A_1, A_2, E\}$): a hypothetical crystal with a single magnetic ion on a $1a$ axis site (site symmetry $C_{3v}$) carries only $\Gamma_{\mathrm{mag}}^{1a} = \Gamma_{\mathrm{axial}} = A_2 \oplus E$, placing the toroidal sector $A_1$ into Class~C; by contrast, the two-ion Fe $2a$ Wyckoff orbit in BiFeO$_3$ (site symmetry $C_3$) yields $\Gamma_{\mathrm{mag}}^{\mathrm{BFO}} = A_1 \oplus A_2 \oplus 2E$ and restores $A_1$ to Class~A. Similarly, a single magnetic ion at an inversion center in a centrosymmetric group (e.g., $1a$ in $D_{3d}$) yields only gerade modes ($A_{2g}\oplus E_g$), placing the entire ungerade support $\mathcal{S}_{\mathrm{ME}}(D_{3d})$ into Class~C ($\mathcal{S}_{\mathrm{act}}^{\mathrm{mat}} = \varnothing$).

Crucially, the kinematic multiplicity $m(\Gamma_L)$ counts how many independent symmetry-adapted physical spin basis mode copies $\mathbf{L}_{\Gamma_L}^{(\mu)}$ ($\mu = 1, \dots, m(\Gamma_L)$) of symmetry $\Gamma_L$ exist on $\mathcal{S}_{\mathrm{mag}}$ ($m(\Gamma_L) = m_{\Gamma_L}^{\mathrm{mat}} = \langle \Gamma_L, \Gamma_{\mathrm{mag}}^{\mathrm{mat}} \rangle$ for real-type irreps, and $m(E) = m_\sigma^{\mathrm{mat}}$ for complex-type 2D real doublets $E = \sigma \oplus \sigma^*$), whereas the constitutive multiplicity $N(\Gamma_L)$ counts how many independent real linear-ME tensor templates are generated once a single condensed mode $\mathbf{L} \sim \Gamma_L$ is selected. By Schur's lemma, every copy $\mathbf{L}_{\Gamma_L}^{(\mu)}$ (or any normal-mode linear combination thereof selected by microscopic exchange interactions) transforms under the same irrep $\Gamma_L$ and therefore shares the identical equivariant intertwiner space $\operatorname{Hom}_G(\Gamma_L, \Gamma_{\mathrm{ME}}^{[2]})$ and tensor templates $\mathbf{C}_{ij}^{(n)}$, differing only in the material-specific coupling amplitudes $\lambda_n$. Throughout this work, we consider magnetic ordering associated with a single selected irreducible normal mode $\mathbf{L} \sim \Gamma_L$ at $\mathbf{k}=0$ (suppressing the copy index $\mu$ unless needed), where $\mathbf{L}$ may have arbitrary dimensionality and an arbitrary order-parameter direction (OPD) within the irrep space; simultaneous condensation of multiple distinct copies lies beyond the present single-mode framework.

\textit{Projection and Invariant Expansion.---} The channel basis is obtained constructively via the Wigner matrix-unit projection operators acting on the nine-dimensional field-dyad space:
\begin{equation}
\mathcal{P}_{k\mu}^{(\Gamma_L)} = \frac{d_L}{|G|} \sum_{g \in G} \left[ D_{k\mu}^{(\Gamma_L)}(g) \right]^* \hat{D}_{\mathrm{ME}}(g),
\label{eq:projector_def}
\end{equation}
where $\hat{D}_{\mathrm{ME}}(g) = R_g \otimes [\det(R_g) R_g]$. (For complex-type real doublets $E = \sigma \oplus \sigma^*$, Eq.~\eqref{eq:projector_def} is applied to the conjugate complex irrep components $\sigma, \sigma^*$ before forming real Cartesian multiplets.) While the trace $\mathcal{P}_{\Gamma_L}^{\mathrm{central}} = \sum_{k=1}^{d_L} \mathcal{P}_{kk}^{(\Gamma_L)}$ is the central idempotent projecting onto the full $\Gamma_L$-isotypic subspace, the matrix-unit operators $\mathcal{P}_{k\mu}^{(\Gamma_L)}$ ($\mathcal{P}_{k\mu}^{(\Gamma_L)}\mathcal{P}_{\nu l}^{(\Gamma_L')} = \delta_{\Gamma_L,\Gamma_L'}\delta_{\mu\nu}\mathcal{P}_{kl}^{(\Gamma_L)}$) constructively extract partner-aligned basis multiplets transforming under the exact same representation matrices $D^{(\Gamma_L)}(g)$ as $\mathbf{L}$, whereas intertwining characterizes the complete channel space (\hyperref[app:category]{Appendix~C}). Eliminating redundancies among the projected dyads yields the linearly independent symmetry-adapted bilinear field multiplets $\mathbf{B}^{(n)}(\mathbf{E},\mathbf{H}) = \sum_{ij} \mathbf{C}_{ij}^{(n)} E_i H_j$, defining geometric tensor templates $C_{ij, k}^{(n)} = \partial^2 B_k^{(n)}/\partial E_i \partial H_j$. Contracting $\mathbf{B}^{(n)}$ with $\mathbf{L}$ constructs the leading-order thermodynamic potential:
\begin{equation}
\begin{aligned}
F_{\mathrm{me}}^{(1)} &= -\sum_{n=1}^N \lambda_n I^{(n)}(\mathbf{L}, \mathbf{E}, \mathbf{H}) \\
&= -\sum_{n=1}^N \lambda_n \sum_{k=1}^{d_L} L_k B_k^{(n)}(\mathbf{E}, \mathbf{H}),
\end{aligned}
\label{eq:F_me_expansion}
\end{equation}
where $\lambda_n$ are material-dependent microscopic coupling coefficients.

\begin{theorem}[Constitutive Generation of Response Tensor]
\label{thm:constitutive}
If a condensed magnetic order parameter $\mathbf{L}$ on $\mathcal{S}_{\mathrm{mag}}$ transforms according to an active irrep $\Gamma_L \in \mathcal{S}_{\mathrm{act}}^{\mathrm{mat}}$ with non-zero constitutive multiplicity $N > 0$ in the parent ME response space, the most general symmetry-allowed leading-order linear magnetoelectric tensor is constitutively generated as:
\begin{equation}
\begin{aligned}
\alpha_{ij}^{(1)}(\mathbf{L}) &= -\left.\frac{\partial^2 F_{\mathrm{me}}^{(1)}}{\partial E_i \partial H_j}\right|_{\mathbf{L}} = \sum_{n=1}^N \lambda_n [\Phi_n(\mathbf{L})]_{ij} \\
&\equiv \sum_{n=1}^N \lambda_n \sum_{k=1}^{d_L} L_k C_{ij, k}^{(n)},
\end{aligned}
\label{eq:tensor_generation}
\end{equation}
where $\{\Phi_n\}_{n=1}^N$ is a chosen basis of equivariant intertwiners in $\operatorname{Hom}_G(\Gamma_L, \Gamma_{\mathrm{ME}}^{[2]})$ with partner-aligned Cartesian matrix templates $C_{ij,k}^{(n)} = [\Phi_n(\hat{\mathbf{e}}_k)]_{ij} = (\mathbf{C}_{ij}^{(n)})_k$.
\end{theorem}

\textit{Geometric Invariance under Order-Parameter Rotation.---}
Because $\mathcal{C}_{\mathrm{mat}} = \sum_{n=1}^N \lambda_n \Phi_n$ is a $G$-equivariant linear map between real orthogonal representations, the symmetric semi-definite operator $\mathcal{C}_{\mathrm{mat}}^{\mathrm{T}}\mathcal{C}_{\mathrm{mat}}$ commutes with $D^{(\Gamma_L)}(g)$ for all $g \in G$. By Schur's lemma over $\mathbb{R}$, $\mathcal{C}_{\mathrm{mat}}^{\mathrm{T}}\mathcal{C}_{\mathrm{mat}} = \kappa I_{d_L}$ ($\kappa \ge 0$) for every physically irreducible real magnetic mode $\Gamma_L$ across all 32 point groups (including complex-type real doublets, whose antisymmetric endomorphism $J$ cannot appear in a symmetric operator), yielding:
\begin{equation}
\left\|\bm{\alpha}^{(1)}(\mathbf{L})\right\|_F^2 = \kappa |\mathbf{L}|^2, \qquad
\bm{\alpha}^{(1)}(\mathbf{L}) : \bm{\alpha}^{(1)}(\mathbf{L}') = \kappa\,\mathbf{L}\cdot\mathbf{L}',
\label{eq:frobenius_invariance}
\end{equation}
where $\|\bm{\alpha}\|_F^2 = \sum_{ij}\alpha_{ij}^2$ and the colon denotes the Frobenius inner product. Thus, for a fixed irreducible mode, fixed mode composition, and fixed coupling coefficients at leading order, rotating $\mathbf{L}$ at constant amplitude $|\mathbf{L}|$ redistributes individual tensor entries while preserving both the total Frobenius norm and the normalized inner product $\hat{\mathbf{L}}\cdot\hat{\mathbf{L}}'$ (and likewise within each multipolar sector separately). In multi-orientation first-principles calculations, Eq.~\eqref{eq:frobenius_invariance} provides a necessary consistency check on the fixed-mode leading-order description; deviations indicate a breakdown of at least one of these assumptions.

\textit{Realization Conditions and Bridge to Downstream MPG Tensor Constraints.---}
To avoid conflating symmetry allowance with physical realization, we summarize the conditions required for an observable leading-order linear ME response in a given material:
\begin{equation}
\begin{aligned}
\text{(i)}\;\; &\Gamma_L \in \Supp\left(\Gamma_{\mathrm{mag}}^{\mathrm{mat}}\right),\\
\text{(ii)}\;\; &\operatorname{Hom}_G(\Gamma_L, V_{\mathrm{ME}}) \neq \{0\},\\
\text{(iii)}\;\; &\mathbf{L}^{(0)} \neq \mathbf{0},\\
\text{(iv)}\;\; &\mathcal{C}_{\mathrm{mat}}(\mathbf{L}^{(0)}) \equiv \sum_{n=1}^N \lambda_n \Phi_n(\mathbf{L}^{(0)}) \neq \mathbf{0}.
\end{aligned}
\label{eq:four_levels}
\end{equation}
Conditions~(i) and (ii) form the two-tier symmetry criterion $\Gamma_L \in \mathcal{S}_{\mathrm{act}}^{\mathrm{mat}}$ (sublattice kinematic availability and parent symmetry compatibility), whereas Condition~(iii) reflects thermodynamic order selection by microscopic energetics. Once $\mathbf{L}^{(0)} \neq \mathbf{0}$ condenses and selects the residual Shubnikov magnetic point group $M_{\mathrm{sub}}(\mathbf{L}^{(0)}) = \operatorname{Stab}_{G\times\Theta_{\mathcal{T}}}(\mathbf{L}^{(0)})$, evaluating the $N$ independent constitutive templates at $\mathbf{L}^{(0)}$ embeds the parent-generated response into the residual MPG-invariant tensor space, $\operatorname{Im}(\operatorname{ev}_{\mathbf{L}^{(0)}}) \subseteq \operatorname{Fix}_{M_{\mathrm{sub}}(\mathbf{L}^{(0)})}(V_{\mathrm{ME}})$ (see \hyperref[app:category]{Appendix~C}). Because the $N$ templates remain linearly independent at any nonzero $\mathbf{L}^{(0)}$ within an irreducible sector $\Gamma_L$, the entire leading-order tensor $\alpha_{ij}^{(1)}(\mathbf{L}^{(0)})$ vanishes in a single-domain state if and only if all $N$ effective channel coefficients vanish simultaneously ($\lambda_1 = \dots = \lambda_N = 0$), whereas inter-channel cancellation can only suppress individual tensor components. Moreover, while the generated space coincides with $\operatorname{Fix}_{M_{\mathrm{sub}}}(V_{\mathrm{ME}})$ for any one-dimensional real sign irrep $\Gamma_L = \chi_L$, for multidimensional irreps it can either span the full residual space or form a strict $N(\Gamma_L)$-dimensional subspace within $\operatorname{Fix}_{M_{\mathrm{sub}}}(V_{\mathrm{ME}})$, enforcing additional leading-order constitutive relations among tensor components. Within the scope of this paper, we focus on these direct $O(L)$ constitutive relations; experimentally observed deviations from them can serve as a diagnostic of higher odd-order terms in $\mathbf{L}$ (such as $L^3 E_i H_j$, which remain linear in the probe fields), external symmetry-breaking biases, or additional condensed magnetic modes.

\subsection{Multipolar Decomposition and Observable Selectivity}

The ambient geometric symmetry governing electromagnetic field dyads is the full orthogonal group $O(3) = SO(3) \times \{E, \mathcal{I}\}$. Because the electric field is polar ($\mathbf{E} \sim D_1^-$) and the magnetic field is axial ($\mathbf{H} \sim D_1^+$), the field product space transforms under $O(3)$ as the parity-twisted representation:
\begin{equation}
D_1^- \otimes D_1^+ \cong D_0^- \oplus D_1^- \oplus D_2^-\!\quad (D_J^- \equiv D^{(J)} \otimes D_0^-),
\label{eq:O3_decomp}
\end{equation}
where $D_0^- \equiv \det_{O(3)}$ denotes odd determinant parity and $D^{(J)}$ is the inversion-even extension of the angular-momentum-$J$ irrep of $SO(3)$ to $O(3)$. Any macroscopic rank-two magnetoelectric tensor $\alpha_{ij}$ admits the established multipolar expansion \cite{SpaldinFiebigMostovoy2008,Hehl2009,WatanabeYanase2018}:
\begin{equation}
\alpha_{ij} = \theta\,\delta_{ij} + \epsilon_{ijk}(T_\alpha)_k + Q_{S,ij},
\label{eq:ME_multipole_decomposition}
\end{equation}
decomposing into the pseudoscalar monopolar trace $\theta = \frac{1}{3}\alpha_{kk}$, the antisymmetric toroidal response vector $(T_\alpha)_k = \frac{1}{2}\epsilon_{ijk}\alpha_{ij}$ (denoting the toroidal component of the ME response tensor rather than a microscopic magnetic toroidal moment), and the symmetric-traceless quadrupole $Q_{S,ij} = \frac{1}{2}(\alpha_{ij}+\alpha_{ji}) - \theta\delta_{ij}$ (Table~\ref{tab:ME_irrep_dictionary}). These macroscopic response sectors transform under the same $O(3)\downarrow G$ representations as the corresponding microscopic position--spin magnetoelectric multipoles ($\int r_i \mu_j(\mathbf{r})\,d^3r$), but no numerical identification between the macroscopic response coefficients and a particular atomic-scale multipole moment is assumed here \cite{SpaldinFiebigMostovoy2008}.

\begin{table}[t]
\caption{Multipolar correspondence of the linear magnetoelectric response.}
\label{tab:ME_irrep_dictionary}
\centering
\small
\setlength{\tabcolsep}{4.5pt}
\renewcommand{\arraystretch}{1.2}
\begin{ruledtabular}
\begin{tabular}{ccccc}
Sector & $O(3)$ & Multipole & $\alpha_{ij}$ Form & Invariant \\
\midrule
Monopolar   & $D_0^-$ & $\theta$         & $\theta\,\delta_{ij}$          & $\theta\,\mathbf{E}\cdot\mathbf{H}$ \\
Toroidal    & $D_1^-$ & $\mathbf{T}_\alpha$ & $\epsilon_{ijk}(T_\alpha)_k$   & $\mathbf{T}_\alpha\cdot(\mathbf{E}\times\mathbf{H})$ \\
Quadrupolar & $D_2^-$ & $\mathbf{Q}_S$   & $Q_{S,ij}$                     & $\mathbf{E}\cdot\mathbf{Q}_S\cdot\mathbf{H}$ \\
\end{tabular}
\end{ruledtabular}
\end{table}

Under subgroup subduction $O(3) \downarrow G$, the continuous determinant representation restricts to the parent pseudoscalar representation: $D_0^- \downarrow_G \;\equiv \Gamma_{\mathrm{ps}}$. Consequently, crystallographic multipolar branchings satisfy:
\begin{equation}
\boxed{
D_J^- \downarrow_G \;\equiv\; \left( D^{(J)} \downarrow_G \right) \otimes \Gamma_{\mathrm{ps}} = \mathcal{F}_{\mathrm{ps}}\left( D^{(J)} \downarrow_G \right)
}.
\label{eq:subduction_universal}
\end{equation}
This reveals the physical nature of each multipolar sector: the isotropic monopole $\theta$ transforms as $\Gamma_{\mathrm{ps}}$, the toroidal response vector $\mathbf{T}_\alpha$ transforms as a polar vector $\Gamma_{\mathrm{pol}}$ coupling to $\mathbf{T}_\alpha \cdot (\mathbf{E}\times\mathbf{H})$, and the quadrupole $\mathbf{Q}_S$ mediates traceless anisotropic diagonal and shear cross-couplings.

An order parameter $\mathbf{L} \sim \Gamma_L$ selectively activates only those multipolar components whose subduced sectors contain $\Gamma_L$:
\begin{equation}
\begin{cases}
\theta \text{ active} \iff \langle \Gamma_L, \Gamma_{\mathrm{ps}} \rangle > 0, \\[2pt]
\mathbf{T}_\alpha \text{ active} \iff \langle \Gamma_L, \Gamma_{\mathrm{pol}} \rangle > 0, \\[2pt]
\mathbf{Q}_S \text{ active} \iff \langle \Gamma_L, D_2^- \downarrow_G \rangle > 0.
\end{cases}
\label{eq:multipolar_activation}
\end{equation}
Crucially, the constitutive channel multiplicity strictly partitions among these multipolar sectors:
\begin{equation}
\boxed{
N = \left\langle \Gamma_L, \Gamma_{\mathrm{ME}}^{[2]} \right\rangle = N_{\mathrm{mono}} + N_{\mathrm{toro}} + N_{\mathrm{quad}}
}.
\label{eq:N_subduction_sum}
\end{equation}
Equation~\eqref{eq:N_subduction_sum} demonstrates that $N$ is not merely an abstract representation multiplicity, but counts the independent symmetry-allowed constitutive response channels generated by the magnetic irrep. A single magnetic sector can simultaneously activate multiple multipolar channels (e.g., $A_{1u}$ in $D_{3d}$ carries both $J=0^-$ and $J=2^-$, giving $N = 1 + 0 + 1 = 2$). Furthermore, the antisymmetric difference:
\begin{equation}
\Delta_{ij} \equiv \alpha_{ij} - \alpha_{ji} = 2\epsilon_{ijk} (T_\alpha)_k
\label{eq:toroidal_probe}
\end{equation}
identically cancels monopolar and quadrupolar contributions, isolating the antisymmetric toroidal ME response channel.

\subsection{Universal Symmetry Mechanisms}

The subduction identity [Eq.~\eqref{eq:subduction_universal}] organizes the 32 crystallographic point groups into three operational families governed by the algebraic character of $\Gamma_{\mathrm{ps}}$:
(i)~\emph{Centrosymmetric Groups (11 groups):} Spatial inversion satisfies $\mathcal{I} \in G$ with $\Gamma_{\mathrm{ps}} = \Gamma_u$. All subduced sectors are strictly inversion-odd ($D_J^- \downarrow_G \;\subset\; \operatorname{Rep}_u(G)$), enforcing an absolute \emph{parity filter}: only ungerade ($u$-type) magnetic modes can generate linear ME response, rendering all gerade ($g$-type) modes dark channels.
(ii)~\emph{Non-Centrosymmetric, Non-Chiral Groups (10 groups):} Inversion is absent, but improper operations ($\sigma$ or $S_n$) define a non-trivial 1D reflection-parity grading $\Gamma_{\mathrm{ps}}$ (e.g., $A_2$ in $C_{3v}$, $A''$ in $C_s$). The $g/u$ parity obstruction lifts, allowing axial-vector sectors to become active when supported, though higher-order rotational symmetries can still enforce dark channels (Table~\ref{tab:me_incomplete_active}).
(iii)~\emph{Chiral Groups (11 groups):} Proper rotations only enforce $\det(R_g) \equiv +1$, collapsing the pseudoscalar twist to the trivial representation ($\Gamma_{\mathrm{ps}} \equiv \Gamma_1$) and reducing response screening directly to polar tensor decomposition.

This operational branching directly controls whether a parent point group saturates its physical representation support, leading to the global 14/18 classification established below.

\section{Materials Case Studies}

To demonstrate how the standardized parent-group protocol operates across active, dark, and multidimensional regimes---serving both as physical case studies and as analytical benchmarks for algorithmic implementation---we apply the same standardized workflow (crystal and Wyckoff input, sublattice mode decomposition, two-tier screening, channel counting and partner alignment, invariant expansion and tensor generation, and residual MPG comparison) to four representative crystals (Table~\ref{tab:case_summary}). Together, their representative branches illustrate four complementary response scenarios: 1D isomorphism ($2\mathrm{D}=2\mathrm{D}$ in $\mathrm{Cr_2O_3}$ $A_{1u}$), symmetry-enforced zero output ($\alpha\text{-}\mathrm{Fe_2O_3}$), strict multidimensional subspace embedding ($3\mathrm{D} \subset 5\mathrm{D}$ at $\phi=\pi/2$ in $\mathrm{BiFeO_3}$ $E$ and $\mathrm{Cr_2O_3}$ $E_u$), and full multidimensional coverage ($2\mathrm{D}=2\mathrm{D}$ in $\mathrm{Cu_2OSeO_3}$ $T$ along $[001]$).

\begin{table*}[t]
\caption{Two-tier symmetry screening and observable linear magnetoelectric geometry across representative materials. The table connects the space group $\mathcal{G}$ and point group $G$, the $\mathbf{k}=0$ sublattice modes $\Gamma_{\mathrm{mag}}^{\mathrm{mat}}$, the active multipolar sectors $\mathcal{S}_{\mathrm{act}}^{\mathrm{mat}}$, the examined magnetic sectors $\Gamma_L$ with their intrinsic MPGs, and the resulting symmetry origin and measurement geometry.}
\label{tab:case_summary}
\centering
\footnotesize
\setlength{\tabcolsep}{3.5pt}
\renewcommand{\arraystretch}{1.20}
\begin{ruledtabular}
\begin{tabular}{p{1.45cm} p{1.40cm} p{2.55cm} p{3.35cm} p{3.20cm} p{4.15cm}}
\RaggedCell{\textbf{Material}} & \RaggedCell{\textbf{Space group $\mathcal{G}$ ($G$)}} & \RaggedCell{\textbf{Sublattice Modes $\Gamma_{\mathrm{mag}}^{\mathrm{mat}}$}} & \RaggedCell{\textbf{Active Set $\mathcal{S}_{\mathrm{act}}^{\mathrm{mat}}$ \& Multipolar Channels}} & \RaggedCell{\textbf{Examined Sectors $\Gamma_L$ \& Intrinsic MPGs}} & \RaggedCell{\textbf{Symmetry Origin \& Measurement Geometry}} \\
\midrule
\RaggedCell{Cr$_2$O$_3$} & \RaggedCell{$R\bar{3}c\,(D_{3d})$} & \RaggedCell{$A_{1g}\oplus A_{2g}\oplus 2E_g \;\oplus$ \newline $A_{1u}\oplus A_{2u}\oplus 2E_u$} & \RaggedCell{$A_{1u}$ (monopole, quad.); \newline $A_{2u}$ (toroidal); \newline $E_u$ (toroidal, $2\times$quad.)} & \RaggedCell{$A_{1u}$ ($\bar{3}'m'$, ground); $2\mathrm{D}=2\mathrm{D}$; \newline $E_u$ (spin-flop, $\phi=\pi/2$); \newline $2/m'$; $3\mathrm{D}\subset 5\mathrm{D}$} & \RaggedCell{Inversion-odd ($+-+-$) order: $A_{1u}$ splits into isotropic/anisotropic channels; in $E_u$, $\mathbf{T}_\alpha \perp \mathbf{L}_\perp$ and $\Delta\mathbf{P}_\perp(\delta H_z) \parallel \mathbf{L}_\perp$.} \\
\RaggedCell{$\alpha$-Fe$_2$O$_3$} & \RaggedCell{$R\bar{3}c\,(D_{3d})$} & \RaggedCell{$A_{1g}\oplus A_{2g}\oplus 2E_g \;\oplus$ \newline $A_{1u}\oplus A_{2u}\oplus 2E_u$} & \RaggedCell{$A_{1u}$ (monopole, quad.); \newline $A_{2u}$ (toroidal); \newline $E_u$ (toroidal, $2\times$quad.)} & \RaggedCell{$E_g$ (\mbox{$T_{\mathrm{M}}\!<\!T\!<\!T_{\mathrm{N}}$}, \mbox{$2'/m'$}); \newline \mbox{$A_{1g}$ ($T < T_{\mathrm{M}}, \bar{3}m$)} \newline \textit{[Class~B: ME-dark]}} & \RaggedCell{Inversion-even ($+--+$) order forbids net bulk $\alpha_{ij}$ at all orders despite weak ferromagnetism; local compensating responses may persist.} \\
\RaggedCell{BiFeO$_3$} & \RaggedCell{$R3c\,(C_{3v})$} & \RaggedCell{$A_1 \oplus A_2 \oplus 2E$} & \RaggedCell{$A_1$ (toroidal); \newline $A_2$ (monopole, quad.); \newline $E$ (toroidal, $2\times$quad.)} & \RaggedCell{$A_1$ ($L_z^{\mathrm{AFM}}, 3m$); \newline $A_2$ ($M_z^{\mathrm{FM}}, 3m'$); \newline $E$ ($\phi=\pi/2, m'$); $3\mathrm{D}\subset 5\mathrm{D}$} & \RaggedCell{Under $\mathbf{H}_\perp$, pure $A_1$ gives $\Delta\mathbf{P}_\perp\!\perp\!\mathbf{H}_\perp$ while $A_2$ gives $\Delta\mathbf{P}_\perp\!\parallel\!\mathbf{H}_\perp$; in $E$, $\mathbf{T}_\alpha\!\parallel\!\mathbf{L}_\perp$ and $\alpha_{zz}^{(1)}=0, \alpha_{xx}^{(1)}+\alpha_{yy}^{(1)}=0$ (can be lifted at $O(L^3)$).} \\
\RaggedCell{Cu$_2$OSeO$_3$} & \RaggedCell{$P2_1 3\,(T)$} & \RaggedCell{$4A \oplus 4E \oplus 12T$} & \RaggedCell{$A$ (monopole); \newline $E$ (quadrupole); \newline $T$ (toroidal, quad.)} & \RaggedCell{$T$ ($\mathbf{H}_0$-polarized, $\mathbf{L}_T = L\hat{\mathbf{n}}$); \newline $2'2'2$ for $[001]$; $2\mathrm{D}=2\mathrm{D}$} & \RaggedCell{$\mathbf{T}_\alpha = \lambda_A L\hat{\mathbf{n}}$; $(\alpha_{ij}\pm\alpha_{ji})/2$ separates symmetric/antisymmetric parts, while $\hat{\mathbf{n}}\parallel\delta\mathbf{H}\parallel[110]$ gives transverse $\Delta\mathbf{P}_{\mathrm{dir}}^{(1)}$ purely from $\lambda_S$.} \\
\end{tabular}
\end{ruledtabular}
\end{table*}

\subsection{\texorpdfstring{Cr$_2$O$_3$ ($D_{3d}$)}{Cr2O3 (D3d)}: Multi-Channel Diagonal Response and Spin-Flop Reorganization}
\textit{How does a single antiferromagnetic order parameter generate multiple independent response channels, and how do they reorganize under spin flop?---}
In corundum Cr$_2$O$_3$ (space group $\mathcal{G} = R\bar{3}c$, point group $G = D_{3d}$), the historical archetype of linear magnetoelectricity \cite{Dzyaloshinskii1960,Astrov1961}, $\Gamma_{\mathrm{ps}} = A_{1u}$. On the four Cr ions of the rhombohedral $4c$ Wyckoff orbit (site symmetry $C_3$, ordered by increasing fractional coordinate $z_1 < z_2 < z_3 < z_4$ along the $[111]$ $c$ axis such that spatial inversion $\mathcal{I}$ interchanges the outer and inner inversion pairs $(1\leftrightarrow 4)$ and $(2\leftrightarrow 3)$ while $C_{2x}$ interchanges $(1\leftrightarrow 2)$ and $(3\leftrightarrow 4)$), the 12-dimensional $\mathbf{k}=0$ sublattice magnetic representation decomposes into parity-even and parity-odd sectors \cite{Liu_mapping}:
\begin{equation}
\begin{aligned}
\Gamma_{\mathrm{mag}}^{\mathrm{Cr_2O_3}} &= (A_{1g}\oplus A_{2g}\oplus 2E_g) \\
&\quad \oplus (A_{1u}\oplus A_{2u}\oplus 2E_u).
\end{aligned}
\label{eq:Cr2O3_mag_rep}
\end{equation}
Meanwhile, decomposing the parent ME response space yields:
\begin{equation}
\Gamma_{\mathrm{ME}}^{D_{3d}} \cong \underbrace{A_{1u}}_{\text{monopole}} \oplus \underbrace{(A_{2u}\oplus E_u)}_{\text{toroidal}} \oplus \underbrace{(A_{1u}\oplus 2E_u)}_{\text{quadrupole}},
\label{eq:D3d_ME}
\end{equation}
with parent support $\mathcal{S}_{\mathrm{ME}}(D_{3d}) = \{A_{1u}, A_{2u}, E_u\}$. Two-tier screening [Eq.~\eqref{eq:two_tier_intersection}] immediately yields $\mathcal{S}_{\mathrm{act}}^{\mathrm{Cr_2O_3}} = \{A_{1u}, A_{2u}, E_u\}$ (Class~A), whereas all inversion-even sublattice modes $\{A_{1g}, A_{2g}, E_g\}$ fall into Class~B (kinematically available on the $4c$ sublattice but net bulk ME-dark).

In the zero-field ground state of Cr$_2$O$_3$, superexchange selects the ($+-+-$) spin combination $\mathbf{L}_{+-+-} = \mathbf{S}_1 - \mathbf{S}_2 + \mathbf{S}_3 - \mathbf{S}_4$ along the $c$ axis. Because $\mathcal{I}$ swaps $(1\leftrightarrow 4)$ and $(2\leftrightarrow 3)$ while preserving axial spin vectors, $\mathcal{I}\mathbf{L}_{+-+-} = -\mathbf{L}_{+-+-}$ (inversion-odd); simultaneously, $C_{2x}$ swaps $(1\leftrightarrow 2)$ and $(3\leftrightarrow 4)$ (site factor $-1$) and flips $S_z \mapsto -S_z$ (spin factor $-1$, giving $+1$ overall under $C_{2x}$ and $+1$ under $C_{3z}$), confirming condensation in the unique $\Gamma_L^{(0)} = A_{1u}$ sector ($m(A_{1u})=1$). Channel counting [Eq.~\eqref{eq:untwisted_channel_count}] yields $N(A_{1u}) = \langle A_{1g}, \Gamma_{\mathrm{pol}}^{\otimes 2} \rangle = 2$, comprising one scalar monopole invariant ($I_{\mathrm{mono}} = \mathbf{E}\cdot\mathbf{H}$) and one diagonal quadrupole invariant ($I_{\mathrm{quad}} = 2E_zH_z - E_xH_x - E_yH_y$) \cite{Hehl2009,WatanabeYanase2018}. Differentiating $F_{\mathrm{me}}^{(1)} = -L_{A_{1u}}(\lambda_{\mathrm{mono}} I_{\mathrm{mono}} + \lambda_{\mathrm{quad}} I_{\mathrm{quad}})$ yields, at leading order $O(L)$, the two independent principal components of residual MPG $\bar{3}'m'$ ($2\mathrm{D}=2\mathrm{D}$ isomorphism):
\begin{equation}
\begin{aligned}
\alpha_{\perp}^{(1)} &= (\lambda_{\mathrm{mono}} - \lambda_{\mathrm{quad}}) L_{A_{1u}}, \\
\alpha_{\parallel}^{(1)} &= (\lambda_{\mathrm{mono}} + 2\lambda_{\mathrm{quad}}) L_{A_{1u}}.
\end{aligned}
\label{eq:Cr2O3_diagonal}
\end{equation}
Inverting Eq.~\eqref{eq:Cr2O3_diagonal} expresses the two symmetry-independent channels directly in terms of measurable tensor components:
\begin{equation}
\begin{aligned}
\frac{2\alpha_\perp^{(1)} + \alpha_\parallel^{(1)}}{3} &= \lambda_{\mathrm{mono}} L_{A_{1u}}, \\
\frac{\alpha_\parallel^{(1)} - \alpha_\perp^{(1)}}{3} &= \lambda_{\mathrm{quad}} L_{A_{1u}}.
\end{aligned}
\label{eq:Cr2O3_inversion}
\end{equation}
Physically, a single antiferromagnetic order parameter drives both isotropic and anisotropic responses whose relative amplitude is not fixed by symmetry; consequently, observing a vanishing individual component (such as $\alpha_\perp^{(1)} \approx 0$ when $\lambda_{\mathrm{mono}} \approx \lambda_{\mathrm{quad}}$) can reflect inter-channel cancellation even when the net magnetoelectric tensor remains nonzero.

When a strong magnetic field applied along the $c$ axis drives a spin-flop transition \cite{OhtaniKohn1984}, the ($+-+-$) antiferromagnetic vector reorients into the basal plane as $\mathbf{L}_\perp = L(\cos\phi, \sin\phi, 0)$ (where $\phi$ is the physical in-plane azimuth of the N\'{e}el vector), condensing into the active $\Gamma_L = E_u$ sector ($N(E_u)=3$, $1E_{\mathrm{tor}} + 2E_{\mathrm{quad}}$). In the standard polar-vector partner basis where $C_{3z}$ acts as $R(2\pi/3)$ and $C_{2x}$ acts as $\operatorname{diag}(+1, -1)$, the physical N\'{e}el components map with a $90^\circ$ rotation to $\mathbf{L}_{E_u} = (L_1, L_2)^{\mathrm{T}} = (L_{+-+-,y}, -L_{+-+-,x})^{\mathrm{T}} = L(\sin\phi, -\cos\phi)^{\mathrm{T}}$ (corresponding to partner angle $\eta = \phi - \pi/2$). Extracting the three partner-aligned field doublets via Eq.~\eqref{eq:projector_def}---the in-plane toroidal doublet $\boldsymbol{\mathcal{T}}_\perp = (E_yH_z-E_zH_y, \, E_zH_x-E_xH_z)^{\mathrm{T}}$ and the two quadrupolar doublets $\boldsymbol{\mathcal{Q}}_\perp^{(1)} = (E_yH_z+E_zH_y, \, -(E_zH_x+E_xH_z))^{\mathrm{T}}$ and $\boldsymbol{\mathcal{Q}}_\perp^{(2)} = (E_xH_x-E_yH_y, \, -(E_xH_y+E_yH_x))^{\mathrm{T}}$---yields:
\begin{equation}
F_{E_u} = -\mathbf{L}_{E_u}\cdot
\left[
\lambda_T\boldsymbol{\mathcal T}_\perp +
\lambda_{Q1}\boldsymbol{\mathcal Q}_\perp^{(1)} +
\lambda_{Q2}\boldsymbol{\mathcal Q}_\perp^{(2)}
\right],
\label{eq:FEu_hematite}
\end{equation}
which generates up to eight nonzero matrix entries controlled by three independent coefficients:
\begin{align}
\alpha_{xx}^{(1)} &= -\alpha_{yy}^{(1)} = \lambda_{Q2}L_1,\quad \alpha_{xy}^{(1)} = \alpha_{yx}^{(1)} = -\lambda_{Q2}L_2,\nonumber\\
\alpha_{yz}^{(1)} &= (\lambda_{Q1}+\lambda_T)L_1,\quad \alpha_{zy}^{(1)} = (\lambda_{Q1}-\lambda_T)L_1,\nonumber\\
\alpha_{zx}^{(1)} &= (-\lambda_{Q1}+\lambda_T)L_2,\quad \alpha_{xz}^{(1)} = (-\lambda_{Q1}-\lambda_T)L_2.
\label{eq:alpha_Eu_hematite}
\end{align}
Substituting $(L_1, L_2) = (L_{+-+-,y}, -L_{+-+-,x})$ immediately reveals two robust directional relations for any in-plane azimuth $\phi$: (i) the toroidal response vector $\mathbf{T}_\alpha = -\lambda_T\,\hat{\mathbf{z}}\times\mathbf{L}_\perp$ stays strictly perpendicular to the physical N\'{e}el vector and rotates rigidly at $90^\circ$ to it, whereas (ii) an axial probe field $\delta H_z$ induces an in-plane polarization $\Delta\mathbf{P}_\perp = (\lambda_{Q1}+\lambda_T)\delta H_z\,\mathbf{L}_\perp$ collinear with $\mathbf{L}_\perp$ (Table~\ref{tab:E_mode_geometry}). Furthermore, the leading-order relations $\alpha_{zz}^{(1)} = 0$ and $\alpha_{xx}^{(1)} + \alpha_{yy}^{(1)} = 0$ hold for any in-plane orientation $\phi$, while the total Frobenius norm $\|\bm{\alpha}^{(1)}(\phi)\|_F^2 = 2L^2(\lambda_T^2 + \lambda_{Q1}^2 + \lambda_{Q2}^2)$ remains invariant under $\phi$ rotation [Eq.~\eqref{eq:frobenius_invariance}]. At the high-symmetry OPD $\mathbf{L}_{E_u} = (L, 0)^{\mathrm{T}}$ ($\phi = \pi/2$, intrinsic MPG $2/m'$, lowered to joint stabilizer $m'$ under $\mathbf{H}_0 \parallel \hat{\mathbf{z}}$), both $2/m'$ and $m'$ share the same 5D allowed tensor space permitting $\alpha_{zz} \neq 0$; thus Eq.~\eqref{eq:alpha_Eu_hematite} isolates a 3D leading-order subspace. This distinction connects directly to spin-flop measurements reporting a finite $\alpha_{zz}$ \cite{OhtaniKohn1984} and to first-principles spin-lattice calculations by Tillack, Yates, and Radaelli \cite{Tillack2016}. They derived an approximate tensor linear in the spin direction [their Eqs.~(4) and (7)], showing that the dominant spin-flop response contains both symmetric and antisymmetric parts, and also analyzed an additional chiral term cubic in the magnetic order. Note that the $\alpha_{zz} = 0.001\,\mathrm{ps\,m^{-1}}$ reported for the $2/m'$ phase in Ref.~\cite{Tillack2016} sits at their stated numerical uncertainty ($\pm 0.001\,\mathrm{ps\,m^{-1}}$) and should not by itself be taken as resolved evidence for a higher-order response.

\subsection{\texorpdfstring{$\alpha$-Fe$_2$O$_3$ ($D_{3d}$)}{alpha-Fe2O3 (D3d)}: Inversion Parity vs.\ Weak Ferromagnetism}
\textit{Why does an isostructural crystal with the same magnetic sites---and even a weak ferromagnetic moment---exhibit zero net bulk linear ME response?---}
Hematite ($\alpha$-Fe$_2$O$_3$) shares the $R\bar{3}c$ ($D_{3d}$) structure, $4c$ Wyckoff orbit, and 12-dimensional sublattice representation [Eq.~\eqref{eq:Cr2O3_mag_rep}] of Cr$_2$O$_3$, and therefore hosts the exact same Class-A active set $\{A_{1u}, A_{2u}, E_u\}$ \cite{Morin1950,lebRun2020,Verbeek2023,Dzyaloshinsky1958,Moriya1960,Jiang2022,GalindezRuales2025}. Unlike Cr$_2$O$_3$, however, exchange interactions in canonical undoped bulk $\alpha$-Fe$_2$O$_3$ stabilize the ($+--+$) spin sequence $\mathbf{L}_{+--+} = \mathbf{S}_1 - \mathbf{S}_2 - \mathbf{S}_3 + \mathbf{S}_4$ on both sides of the Morin transition ($T_{\mathrm{M}} \approx 260\text{ K}$, $T_{\mathrm{N}} \approx 955\text{ K}$) \cite{Morin1950,Dzyaloshinsky1958}. Because spatial inversion interchanges $(1\leftrightarrow 4)$ and $(2\leftrightarrow 3)$ while preserving axial spin vectors, this arrangement satisfies $\mathcal{I}\mathbf{L}_{+--+} = +\mathbf{L}_{+--+}$, opposite in parity to $\mathcal{I}\mathbf{L}_{+-+-} = -\mathbf{L}_{+-+-}$ in Cr$_2$O$_3$.

Above $T_{\mathrm{M}}$ ($T_{\mathrm{M}} < T < T_{\mathrm{N}}$), the basal-plane antiferromagnetic vector $\mathbf{L}_{+--+} \perp \hat{\mathbf{z}}$ condenses in the inversion-even $\Gamma_L^{(0)} = E_g$ sector (intrinsic MPG $2'/m'$ for a high-symmetry basal-plane OPD \cite{Dzyaloshinsky1958,GalindezRuales2025}). The net in-plane weak-ferromagnetic moment $\mathbf{M}_{\mathrm{WF}} = \sum_{a=1}^4 \mathbf{S}_a$ spans the second $E_g$ copy, admitting the bilinear Dzyaloshinskii--Moriya invariant $D(L_{+--+,x}M_{\mathrm{WF},y} - L_{+--+,y}M_{\mathrm{WF},x})$ responsible for weak-ferromagnetic canting \cite{Dzyaloshinsky1958,Moriya1960}. Below $T_{\mathrm{M}}$, the spins reorient along the $c$ axis while keeping the ($+--+$) sequence, condensing into $\Gamma_L^{(0)} = A_{1g}$ (intrinsic Type-I MPG $\bar{3}m$) \cite{Morin1950,lebRun2020,Dzyaloshinsky1958,Jiang2022,GalindezRuales2025}.

Both canonical bulk phases of ideal $\alpha$-Fe$_2$O$_3$ ($E_g$ and $A_{1g}$) fall into Class~B and preserve spatial inversion $\mathcal{I}$. Because $\mathcal{I}: \bm{\alpha} \mapsto -\bm{\alpha}$, the net bulk static linear ME tensor vanishes identically ($\alpha_{ij} \equiv 0$) at every order in the magnetic amplitude on both sides of $T_{\mathrm{M}}$---a stronger obstruction than a merely vanishing leading-order channel. This comparison yields a clear materials-design principle: \emph{weak ferromagnetism is neither a necessary nor a sufficient condition for a net bulk linear ME response}. Simply rotating the N\'{e}el vector or enhancing weak-ferromagnetic canting within the inversion-even ($+--+$) manifold cannot unlock net bulk linear magnetoelectricity; doing so requires either switching the spin arrangement to an inversion-odd sector or introducing inversion-breaking perturbations (such as heterointerfaces, electric-field biases, or odd-parity polar distortions $Q_u$ enabling mixed couplings $Q_u \mathbf{L}_g \mathbf{E}\mathbf{H}$). Meanwhile, the vanishing net bulk tensor does not exclude mutually compensating site-local (anti-magnetoelectric) linear responses on individual Fe sites lacking local inversion symmetry, as demonstrated for the low-temperature easy-axis phase \cite{Verbeek2023}; both $E_g$ and $A_{1g}$ modes are also symmetry-allowed in the inversion-even third-rank piezomagnetic response space $\Gamma_{\mathrm{PM}}^{[3]}$ (with the low-temperature phase providing a classic example of piezomagnetism \cite{Dzyaloshinskii1958_PM}; see Part~III \cite{Liu_piezo}).

\textit{Symmetry-Predicted Purely Toroidal Candidate on the Corundum $4c$ Sublattice.---}
Beyond explaining the contrast between $\mathrm{Cr_2O_3}$ ($A_{1u}$) and $\alpha\text{-}\mathrm{Fe_2O_3}$ ($A_{1g}$), Eq.~\eqref{eq:Cr2O3_mag_rep} also identifies a third collinear $c$-axis antiferromagnetic candidate on the exact same $R\bar{3}c$ $4c$ sublattice: the ($++--$) spin sequence $\mathbf{L}_{++--} = \mathbf{S}_1 + \mathbf{S}_2 - \mathbf{S}_3 - \mathbf{S}_4$ along $c$. Under $\mathcal{I}$ [$(1\leftrightarrow 4, 2\leftrightarrow 3)$], $\mathbf{L}_{++--}$ reverses sign; under $C_{2x}$ [$(1\leftrightarrow 2, 3\leftrightarrow 4)$], the site permutation is even ($+1$) while $S_z \mapsto -S_z$ gives $-1$, placing $L_{++--,z}$ uniquely in the Class-A $A_{2u}$ sector (intrinsic MPG $D_{3d}(C_{3v}) = \bar{3}'m$, $N(A_{2u})=1$). The three collinear corundum spin sequences have been compared in earlier electronic-structure studies \cite{Moore2007}; here we establish their contrasting homogeneous linear-ME fingerprints. If this ordering is stabilized while retaining the assumed $R\bar{3}c$ parent symmetry and equivalent $4c$ magnetic orbit, it is predicted to exhibit a purely antisymmetric axial toroidal response, $\alpha_{xy}^{(1)} = -\alpha_{yx}^{(1)} = \lambda_T L_{++--,z}$ (all other entries zero), inducing a strictly transverse polarization $\Delta\mathbf{P}_\perp \perp \mathbf{H}_\perp$. Because this 1D mode saturates $\operatorname{Fix}_{\bar{3}'m}(V_{\mathrm{ME}})$, the purely antisymmetric tensor form is fixed exactly by the residual $\bar{3}'m$ symmetry, whereas its amplitude remains a microscopic material parameter. Thus, without altering the crystal structure or magnetic Wyckoff orbit, switching the collinear $4c$ spin sequence selects among diagonal ($A_{1u}$), globally dark ($A_{1g}$), and purely antisymmetric ($A_{2u}$) linear ME responses.

\subsection{\texorpdfstring{BiFeO$_3$ ($C_{3v}$)}{BiFeO3 (C3v)}: Polarization Geometry and In-Plane Angular Locking}
\textit{How do distinct magnetic modes dictate the direction of the field-induced polarization, and how does the tensor rotate with an in-plane N\'{e}el vector?---}
In multiferroic BiFeO$_3$ (space group $\mathcal{G} = R3c$, polar point group $G = C_{3v}$) \cite{CatalanScott2009,Wang2003,Popov1994,Gareeva2022,WojdelIniguez2009}, the absence of spatial inversion lifts the $g/u$ parity filter. On the two Fe ions $(1,2)$ of the $2a$ Wyckoff orbit (site symmetry $C_3$, interchanged by the $c$-glide reflection $\{\sigma_v \vert \bm{\tau}_c\}$), combining the site-permutation representation $\Gamma_{\mathrm{perm}}^{2a} = A_1(1,1) \oplus A_2(1,-1)$ with $\Gamma_{\mathrm{axial}} = A_2(S_z) \oplus E(S_x, S_y)$ yields:
\begin{equation}
\Gamma_{\mathrm{mag}}^{\mathrm{BFO}} = A_1 \oplus A_2 \oplus 2E.
\label{eq:BFO_mag_rep}
\end{equation}
With $\Gamma_{\mathrm{ps}} = A_2$, the parent ME response space decomposes as:
\begin{equation}
\Gamma_{\mathrm{ME}}^{C_{3v}} \cong \underbrace{A_2}_{\text{monopole}} \oplus \underbrace{(A_1\oplus E)}_{\text{toroidal}} \oplus \underbrace{(A_2\oplus 2E)}_{\text{quadrupole}}.
\label{eq:C3v_ME}
\end{equation}
Because $C_{3v}$ is ME-complete, every $\mathbf{k}=0$ mode on the Fe $2a$ sublattice belongs to Class~A ($\mathcal{S}_{\mathrm{act}}^{\mathrm{BFO}} = \{A_1, A_2, E\}$).

Evaluating the $c$-glide action on the two axial spin combinations immediately produces a sharp geometric distinction between the two pure 1D candidate modes under an in-plane probe magnetic field $\mathbf{H}_\perp = (H_x, H_y, 0)$:
\begin{itemize}
\item \textit{Out-of-plane G-type antiferromagnetic mode} $L_z = S_{1z} - S_{2z}$: the site-exchange factor ($-1$) and axial-spin reflection factor ($-1$) multiply to $+1$, placing $L_z$ in $A_1$ (intrinsic Type-I MPG $3m$). Its unique toroidal channel $F_{A_1} = -\lambda_T L_z (E_x H_y - E_y H_x)$ yields $\alpha_{xy}^{(1)} = -\alpha_{yx}^{(1)} = \lambda_T L_z$ (consistent with Eq.~(4) of Ref.~\cite{Gareeva2022}, crediting Ref.~\cite{Popov1994}), inducing a strictly \emph{transverse} polarization change $\Delta\mathbf{P}_\perp = \lambda_T L_z (H_y, -H_x, 0) \perp \mathbf{H}_\perp$.
\item \textit{Out-of-plane ferromagnetic mode} $M_z = S_{1z} + S_{2z}$: transforming as $A_2$ (intrinsic Type-III MPG $3m'$), it activates two diagonal channels ($\alpha_{zz}^{(1)} = \lambda_\parallel M_z$, $\alpha_{xx}^{(1)} = \alpha_{yy}^{(1)} = \lambda_\perp M_z$), inducing a strictly \emph{collinear} polarization change $\Delta\mathbf{P}_\perp = \lambda_\perp M_z (H_x, H_y, 0) \parallel \mathbf{H}_\perp$.
\end{itemize}
Thus, even without knowing the microscopic coupling amplitudes, comparing transverse versus collinear field-induced $\Delta\mathbf{P}_\perp$ (distinct from the spontaneous ferroelectric polarization along $z$) cleanly separates these two pure axial candidate modes.

When the long-period ($\sim 62\text{ nm}$) cycloid of bulk BiFeO$_3$ \cite{CatalanScott2009} is suppressed \cite{Popov1994,Wang2003}, homogeneous in-plane branches reside within the $m(E)=2$ copies of the unperturbed $R3c$ parent (an epitaxial distortion or bias field that lowers this parent symmetry can introduce additional terms). Because the $c$-glide interchanges sites $1\leftrightarrow 2$, in the standard partner basis where $C_{3z}$ acts as $R(2\pi/3)$ and $\sigma_{vy}$ ($y \to -y$) acts as $\operatorname{diag}(-1, +1)$, the physical in-plane antiferromagnetic ($\mathbf{L}_\perp = \mathbf{S}_{1\perp} - \mathbf{S}_{2\perp} = L(\cos\phi, \sin\phi, 0)$) and weak-ferromagnetic ($\mathbf{M}_\perp = \mathbf{S}_{1\perp} + \mathbf{S}_{2\perp}$) doublets align as $\mathbf{L}_E^{\mathrm{AFM}} = (L_y, -L_x)^{\mathrm{T}} = L(\sin\phi, -\cos\phi)^{\mathrm{T}}$ and $\mathbf{M}_E^{\mathrm{FM}} = (M_x, M_y)^{\mathrm{T}}$. Any condensed $E$ normal mode $\mathbf{L}_E = (L_1, L_2)^{\mathrm{T}}$ (including a Dzyaloshinskii--Moriya-locked superposition) activates $N(E)=3$ partner-aligned field doublets $\boldsymbol{\mathcal{B}}_E^{(1)} = (E_zH_x, \, E_zH_y)^{\mathrm{T}}$, $\boldsymbol{\mathcal{B}}_E^{(2)} = (E_xH_z, \, E_yH_z)^{\mathrm{T}}$, and $\boldsymbol{\mathcal{B}}_E^{(3)} = (E_xH_x-E_yH_y, \, -(E_xH_y+E_yH_x))^{\mathrm{T}}$:
\begin{equation}
F_E = -\mathbf{L}_E \cdot \left[ \lambda_1 \boldsymbol{\mathcal{B}}_E^{(1)} + \lambda_2 \boldsymbol{\mathcal{B}}_E^{(2)} + \lambda_3 \boldsymbol{\mathcal{B}}_E^{(3)} \right].
\label{eq:FE_BFO}
\end{equation}
For the pure in-plane antiferromagnetic mode parameterized by its physical N\'{e}el azimuth $\phi$, Eq.~\eqref{eq:FE_BFO} directly yields the angle-resolved leading-order tensor:
\begin{align}
(\alpha_{zx}^{(1)}, \alpha_{zy}^{(1)}) &= \lambda_1 L (\sin\phi, -\cos\phi), \nonumber\\
(\alpha_{xz}^{(1)}, \alpha_{yz}^{(1)}) &= \lambda_2 L (\sin\phi, -\cos\phi), \nonumber\\
\alpha_{xx}^{(1)} = -\alpha_{yy}^{(1)} &= \lambda_3 L \sin\phi, \quad \alpha_{xy}^{(1)} = \alpha_{yx}^{(1)} = \lambda_3 L \cos\phi,
\label{eq:BFO_E_tensor}
\end{align}
with $\alpha_{zz}^{(1)} = 0$, $\alpha_{xx}^{(1)} + \alpha_{yy}^{(1)} = 0$, and constant Frobenius norm $\|\bm{\alpha}^{(1)}(\phi)\|_F^2 = L^2(\lambda_1^2 + \lambda_2^2 + 2\lambda_3^2)$ identically for all $\phi$.

\begin{table}[t]
\caption{Opposite leading-order response geometries of pure in-plane antiferromagnetic doublets in $\mathrm{Cr_2O_3}$ ($E_u$) and $\mathrm{BiFeO_3}$ ($E$) for a physical N\'{e}el vector $\mathbf{L}_\perp = L(\cos\phi, \sin\phi, 0)$ (partner basis $(L_1,L_2) = (L_y, -L_x)$).}
\label{tab:E_mode_geometry}
\centering
\footnotesize
\setlength{\tabcolsep}{4pt}
\renewcommand{\arraystretch}{1.22}
\begin{ruledtabular}
\begin{tabular}{lcc}
Observable relation & $\mathrm{Cr_2O_3}$ ($E_u$) & $\mathrm{BiFeO_3}$ ($E$) \\
\midrule
Toroidal vector $\mathbf{T}_\alpha$ & $-\lambda_T\,\hat{\mathbf{z}}\times\mathbf{L}_\perp$ & $\frac{1}{2}(\lambda_1-\lambda_2)\mathbf{L}_\perp$ \\
Relation to $\mathbf{L}_\perp$ & Perpendicular ($90^\circ$) & Collinear \\
$\Delta\mathbf{P}_\perp$ under $\delta H_z$ & $(\lambda_{Q1}+\lambda_T)\delta H_z\,\mathbf{L}_\perp$ & $-\lambda_2\delta H_z\,\hat{\mathbf{z}}\times\mathbf{L}_\perp$ \\
Relation to $\mathbf{L}_\perp$ & Collinear & Perpendicular ($90^\circ$) \\
\end{tabular}
\end{ruledtabular}
\end{table}

Equation~\eqref{eq:BFO_E_tensor} yields three direct physical consequences. First, contrasting Eq.~\eqref{eq:BFO_E_tensor} with Eq.~\eqref{eq:alpha_Eu_hematite} for Cr$_2$O$_3$ reveals opposite response geometries for the two in-plane antiferromagnetic doublets (Table~\ref{tab:E_mode_geometry}): in BiFeO$_3$, the toroidal vector $\mathbf{T}_\alpha = \frac{1}{2}(\lambda_1 - \lambda_2)\mathbf{L}_\perp$ is \emph{collinear} with the physical N\'{e}el vector $\mathbf{L}_\perp$, whereas an axial probe field $\delta H_z$ induces a \emph{perpendicular} in-plane polarization $\Delta\mathbf{P}_\perp = -\lambda_2 \delta H_z\,\hat{\mathbf{z}}\times\mathbf{L}_\perp$.

Second, at the high-symmetry orientation $\phi = \pi/2$ ($\mathbf{L}_E = (L, 0)^{\mathrm{T}}$, intrinsic monoclinic MPG $m'$), comparing the residual MPG-allowed tensor with the leading-order $E$-mode template makes the $3\mathrm{D} \subset 5\mathrm{D}$ strict subspace explicit:
\begin{equation}
\bm{\alpha}_{m'} = \begin{pmatrix} a & 0 & b \\ 0 & c & 0 \\ d & 0 & e \end{pmatrix}, \qquad
\bm{\alpha}_E^{(1)} = L \begin{pmatrix} \lambda_3 & 0 & \lambda_2 \\ 0 & -\lambda_3 & 0 \\ \lambda_1 & 0 & 0 \end{pmatrix},
\label{eq:BFO_5D_3D}
\end{equation}
reproducing the known leading-order relations $\alpha_{zz}^{(1)} = 0$ and $\alpha_{xx}^{(1)} + \alpha_{yy}^{(1)} = 0$ (consistent with the phenomenological tensor in Eq.~(4) of Ref.~\cite{Gareeva2022}, crediting Ref.~\cite{Popov1994}, up to an in-plane coordinate rotation and coefficient convention). The two missing diagonal directions are not forbidden by the residual $m'$ symmetry; rather, in a pure, analytic single-mode expansion they can first appear at third order in $\mathbf{L}_E$ via the cubic $A_2$ polynomial $f_3(\mathbf{L}_E) = L_1(L_1^2 - 3L_2^2) = -L^3 \sin 3\phi$ (or via an admixed $M_z \sim A_2$ component). Coupling $f_3$ to the two $A_2$ diagonal field bilinears ($E_x H_x + E_y H_y$ and $E_z H_z$) permits the cubic correction
\begin{equation}
\delta\bm{\alpha}^{(3)} = f_3(\mathbf{L}_E)\,\operatorname{diag}(c_\perp, c_\perp, c_\parallel),
\label{eq:BFO_cubic}
\end{equation}
which at $\phi = \pi/2$ ($f_3 = L^3$) can supply the remaining two independent diagonal parameters of $\bm{\alpha}_{m'}$ without changing the residual $m'$ symmetry. A threefold angular harmonic ($\propto L^3 \sin 3\phi$) in $\alpha_{zz}$ and $\alpha_{xx}+\alpha_{yy}$ is therefore allowed if $c_\parallel, c_\perp \neq 0$; symmetry determines its angular form, not its magnitude or experimental visibility.

Third, at leading order, Eq.~\eqref{eq:BFO_E_tensor} defines a quantitative angular-calibration protocol: for a homogeneous $E$-mode branch with controlled amplitude $L$ and mode composition, measuring three independent tensor components at a single reference angle (for example, $\alpha_{zx}^{(1)}$, $\alpha_{xz}^{(1)}$, and $\alpha_{xx}^{(1)}$ at $\phi = \pi/2$) calibrates $\lambda_1 L, \lambda_2 L, \lambda_3 L$ and predicts the leading-order angular dependence of all eight in-plane and cross-plane components as $\phi$ rotates.

\subsection{\texorpdfstring{Cu$_2$OSeO$_3$ ($T$)}{Cu2OSeO3 (T)}: Three-Dimensional Orientation, Probe Selectivity, and Magnetic Pinning}
\textit{How does a three-dimensional magnetic order parameter steer the response tensor, and how do probe geometry and magnetic pinning govern channel separation?---}
Cu$_2$OSeO$_3$ (space group $\mathcal{G} = P2_13$, chiral cubic point group $G = T$) \cite{Bos2008,Seki2012,Belesi2012} has 16 Cu ions on the $4a$ and $12b$ Wyckoff orbits, yielding:
\begin{equation}
\Gamma_{\mathrm{mag}}^{\mathrm{Cu}} = 4A \oplus 4E \oplus 12T,
\label{eq:Cu2OSeO3_mag_rep}
\end{equation}
while $\Gamma_{\mathrm{ps}} \equiv A$ gives the parent response decomposition:
\begin{equation}
\begin{aligned}
\Gamma_{\mathrm{ME}}^{[2]}(T) &= T \otimes T \cong \underbrace{A}_{\text{monopole}} \oplus \underbrace{T_A}_{\text{toroidal}} \oplus \underbrace{(E\oplus T_S)}_{\text{quadrupole}} \\
&\cong A \oplus E \oplus 2T.
\end{aligned}
\label{eq:T_ME_decomp}
\end{equation}
Both the sublattice and parent supports saturate $\mathrm{Irr}_{\mathbb{R}}(T) = \{A, E, T\}$, placing every $\mathbf{k}=0$ sublattice mode in Class~A and leaving zero dark channels.

While the zero-field ground state is a finite-$\mathbf{k}$ helimagnet \cite{Seki2012}, a static biasing magnetic field $\mathbf{H}_0$ above the critical threshold stabilizes a homogeneous $\mathbf{k}=0$ collinear ferrimagnetic normal mode $\mathbf{L}_T = L\hat{\mathbf{n}}$ in the real-type $T$ sector ($N(T)=2$) \cite{Bos2008,Seki2012,Belesi2012}, where $\hat{\mathbf{n}} = (n_x, n_y, n_z) = (\sin\vartheta\cos\varphi, \sin\vartheta\sin\varphi, \cos\vartheta)$ specifies the magnetic orientation in spherical angles $(\vartheta, \varphi)$ (using $\vartheta$ to avoid confusion with the monopolar trace $\theta$). Within this state, the leading-order direct constitutive contribution $\alpha_{\mathrm{dir},ij}^{(1)}(\mathbf{L}_T)$ couples $\mathbf{L}_T$ to the toroidal field vector $\boldsymbol{\mathcal{T}} = \mathbf{E}\times\delta\mathbf{H}$ and the symmetric off-diagonal quadrupolar field triplet $\boldsymbol{\mathcal{Q}}_S = (E_y\delta H_z+E_z\delta H_y, \, E_z\delta H_x+E_x\delta H_z, \, E_x\delta H_y+E_y\delta H_x)^{\mathrm{T}}$:
\begin{equation}
F_{\mathrm{me}}^{(T)} = -\lambda_A\,\mathbf{L}_T\cdot\boldsymbol{\mathcal{T}} - \lambda_S\,\mathbf{L}_T\cdot\boldsymbol{\mathcal{Q}}_S.
\label{eq:Cu2OSeO3_free_energy}
\end{equation}
Belesi \textit{et al.} \cite{Belesi2012} previously used the $P2_13$ symmetry generators and time reversal to construct a parent-symmetry Landau--Ginzburg theory with two homogeneous terms linear in polarization and bilinear in sublattice magnetic variables [their Eqs.~(5)--(8)], separating symmetric-exchange and antisymmetric Dzyaloshinskii--Moriya combinations (with the symmetric term dominating the uniform polarization in the long-wavelength conical state). Here, Eq.~\eqref{eq:Cu2OSeO3_free_energy} integrates the complete pair of direct macroscopic $O(LE\delta H)$ response templates for a selected homogeneous $T$ mode into our universal 32-point-group channel-counting and tensor-generation scheme.

In fixed cubic crystal axes, Eq.~\eqref{eq:Cu2OSeO3_free_energy} generates six off-diagonal elements with vanishing diagonal entries ($\alpha_{\mathrm{dir},xx}^{(1)}=\alpha_{\mathrm{dir},yy}^{(1)}=\alpha_{\mathrm{dir},zz}^{(1)}=0$):
\begin{equation}
\begin{aligned}
\alpha_{\mathrm{dir},yz}^{(1)} &= (\lambda_S + \lambda_A)L n_x, \quad \alpha_{\mathrm{dir},zy}^{(1)} = (\lambda_S - \lambda_A)L n_x, \\
\alpha_{\mathrm{dir},zx}^{(1)} &= (\lambda_S + \lambda_A)L n_y, \quad \alpha_{\mathrm{dir},xz}^{(1)} = (\lambda_S - \lambda_A)L n_y, \\
\alpha_{\mathrm{dir},xy}^{(1)} &= (\lambda_S + \lambda_A)L n_z, \quad \alpha_{\mathrm{dir},yx}^{(1)} = (\lambda_S - \lambda_A)L n_z,
\end{aligned}
\label{eq:Cu2OSeO3_tensor_3D}
\end{equation}
so that the toroidal response vector $\mathbf{T}_\alpha = \lambda_A L\hat{\mathbf{n}}$ always tracks the ferrimagnetic order direction $\hat{\mathbf{n}}$, and $\|\bm{\alpha}_{\mathrm{dir}}^{(1)}(\hat{\mathbf{n}})\|_F^2 = 2L^2(\lambda_S^2 + \lambda_A^2)$ is independent of $\hat{\mathbf{n}}$ [Eq.~\eqref{eq:frobenius_invariance}]. For $\hat{\mathbf{n}} \parallel [001]$ (isotropy subgroup $2'2'2$), $N(T) = 2 = \dim\operatorname{Fix}_{2'2'2}(V_{\mathrm{ME}})$: the residual MPG permits exactly two independent components ($\alpha_{xy}, \alpha_{yx}$), which are fully spanned by the two parent-generated channels ($2\mathrm{D} = 2\mathrm{D}$):
\begin{equation}
\begin{aligned}
\frac{\alpha_{\mathrm{dir},xy}^{(1)} + \alpha_{\mathrm{dir},yx}^{(1)}}{2} &= \lambda_S L, \\
\frac{\alpha_{\mathrm{dir},xy}^{(1)} - \alpha_{\mathrm{dir},yx}^{(1)}}{2} &= \lambda_A L = (T_\alpha)_z.
\end{aligned}
\label{eq:Cu_channel_sep}
\end{equation}
More broadly, as $\hat{\mathbf{n}}$ rotates across $[001]$, $[111]$, $[110]$, and a general orientation, the intrinsic residual MPG ($2'2'2$, $3$, $2'$, and $1$) permits $2$, $3$, $4$, and $9$ independent tensor parameters respectively, whereas the entire three-dimensional leading-order family in Eq.~\eqref{eq:Cu2OSeO3_tensor_3D} remains locked to the same two parent couplings $(\lambda_S, \lambda_A)$. Contrasting the full $2\mathrm{D}=2\mathrm{D}$ coverage at $[001]$ with the strict $3\mathrm{D} \subset 5\mathrm{D}$ embedding in $\mathrm{BiFeO_3}$ ($m'$)---and with the $2\mathrm{D} \subset 3\mathrm{D}, 4\mathrm{D}, 9\mathrm{D}$ embeddings at other $\mathrm{Cu_2OSeO_3}$ orientations---demonstrates both that multidimensional irreps do not automatically produce strict subspaces and that parent generation links all orientations into a single parameter-sparse constitutive tensor family.

Equation~\eqref{eq:Cu2OSeO3_tensor_3D} highlights two practical measurement insights. First, \emph{transverse polarization in a single measurement geometry does not by itself prove a toroidal channel}. Because the antisymmetric polarization contribution is $\Delta\mathbf{P}_A = \lambda_A L\,\delta\mathbf{H}\times\hat{\mathbf{n}}$, applying a longitudinal probe field $\delta\mathbf{H} = h\hat{\mathbf{n}}$ along the magnetic orientation identically extinguishes the direct toroidal term, leaving only the symmetric quadrupolar response:
\begin{equation}
\Delta\mathbf{P}_{\mathrm{dir}}^{(1)}(\delta\mathbf{H} = h\hat{\mathbf{n}}) = 2\lambda_S L h\,(n_y n_z, \, n_z n_x, \, n_x n_y).
\label{eq:Cu_longitudinal_probe}
\end{equation}
Consistent with the directional polarizations observed in $\mathrm{Cu_2OSeO_3}$ \cite{Seki2012,Belesi2012}, a longitudinal probe yields zero direct response along $[001]$, a strictly \emph{transverse} polarization $\Delta\mathbf{P}_{\mathrm{dir}}^{(1)} = \lambda_S L h\,\hat{\mathbf{z}}$ along $[110]$ (arising purely from the symmetric quadrupole $\lambda_S$), and a \emph{collinear} polarization $\Delta\mathbf{P}_{\mathrm{dir}}^{(1)} = (2/\sqrt{3})\lambda_S L h\,\hat{\mathbf{n}}$ along $[111]$. Unambiguously isolating the toroidal channel therefore requires comparing $\alpha_{ij}$ and $\alpha_{ji}$ rather than relying on transverse polarization alone. Second, \emph{fixed-mode direct response must be distinguished from total differential response in a biasing field}: in addition to the direct $O(LEH)$ constitutive contribution $\alpha_{\mathrm{dir},ij}^{(1)} = -\partial^2 F_{\mathrm{me}}^{(T)}/(\partial E_i \partial(\delta H_j))\vert_{\mathbf{L}_T}$, a total differential measurement includes order-parameter relaxation $(\partial P_i/\partial L_a)(\partial L_a/\partial H_j)$ and any paramagnetic bias background allowed by $T$, such as $F_{\mathrm{bg}} = -\beta(E_x H_y H_z + E_y H_z H_x + E_z H_x H_y)$. Under $\mathbf{H}_0 \parallel [001]$, $F_{\mathrm{bg}}$ contributes $\alpha_{xy}^{\mathrm{bg}} = \alpha_{yx}^{\mathrm{bg}} = \beta H_0$, which enters the symmetric combination $(\alpha_{xy}+\alpha_{yx})/2$ alongside $\lambda_S L$ but cancels in the antisymmetric difference $(\alpha_{xy}-\alpha_{yx})/2$. Meanwhile, linearizing $\mathbf{P}_A = \lambda_A \mathbf{H}\times\mathbf{L}$ gives $\delta\mathbf{P}_A = \lambda_A(\delta\mathbf{H}\times\mathbf{L}_0 + \mathbf{H}_0\times\delta\mathbf{L})$, which recovers the direct term when magnetic anisotropy pins $\mathbf{L}$ ($\delta\mathbf{L} \approx \mathbf{0}$) but cancels in the ideal free-tracking limit ($\mathbf{H}\times\mathbf{L}=0$). Magnetic anisotropy and mode relaxation must therefore be accounted for alongside symmetry-allowed templates when interpreting differential ME measurements in biasing fields \cite{Seki2012}.

\section{Discussion}

\subsection{Subgroup Heredity of ME Completeness and Non-Gray Activation}

Evaluating the representation support $\Supp(\Gamma_{\mathrm{ME}}^{[2]})$ across all 32 crystallographic point groups reveals a fundamental partition of parent crystal symmetries. Formally, a parent spatial point group $G$ is defined to be \emph{magnetoelectric-complete} (ME-complete) if and only if its linear ME response representation contains every irreducible representation of $G$:
\begin{equation}
\boxed{
G \text{ is ME-complete} \iff \Supp(\Gamma_{\mathrm{ME}}^{[2]}) = \mathrm{Irr}(G)
}.
\label{eq:me_complete_def}
\end{equation}
Physically, ME completeness ensures that the parent response space contains at least one leading-order constitutive channel for every irreducible magnetic sector, guaranteeing that no representation-level dark magnetic sector exists. Globally, exactly 14 parent point groups are ME-complete, while the remaining 18 are ME-incomplete (\hyperref[app:tables]{Appendix~B}, Tables~\ref{tab:me_complete_induced}--\ref{tab:me_inactive_with_irreps}).

Crucially, ME completeness is a universal parent-group property defined on $\mathrm{Irr}(G)$ independently of specific Wyckoff occupations. Within the two-tier screening framework [Eq.~\eqref{eq:two_tier_intersection}], an ME-complete parent group ($\mathcal{S}_{\mathrm{ME}}(G) = \mathrm{Irr}(G)$) guarantees $\mathcal{S}_{\mathrm{act}}^{\mathrm{mat}} = \mathcal{S}_{\mathrm{mag}}^{\mathrm{mat}} \cap \mathrm{Irr}(G) = \mathcal{S}_{\mathrm{mag}}^{\mathrm{mat}}$, meaning that parent response symmetry imposes no additional filtering and every magnetic mode hosted by the sublattice is automatically ME-active. Conversely, for the 18 ME-incomplete parent groups ($\mathcal{S}_{\mathrm{ME}}(G) \subsetneq \mathrm{Irr}(G)$), parent symmetry can filter the sublattice modes, $\mathcal{S}_{\mathrm{act}}^{\mathrm{mat}} \subseteq \mathcal{S}_{\mathrm{mag}}^{\mathrm{mat}}$, projecting any mode in $\mathcal{S}_{\mathrm{mag}}^{\mathrm{mat}} \setminus \mathcal{S}_{\mathrm{ME}}(G)$ into a leading-order dark channel.

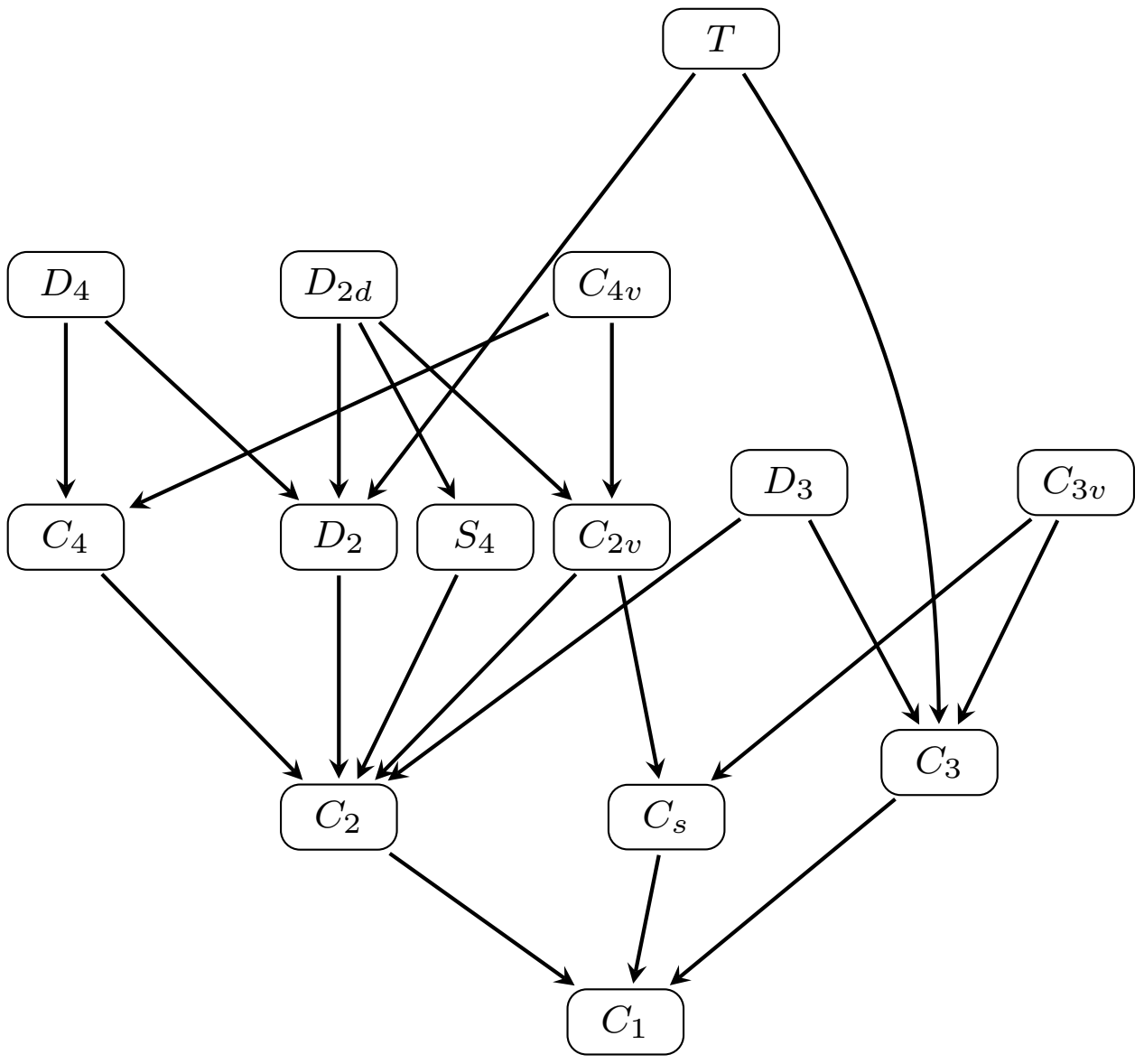
\begin{figure}[t]
\centering
\begin{tikzpicture}[
  node distance=1.3cm and 1.3cm,
  every node/.style={draw, rounded corners, fill=white, inner sep=3.5pt, minimum width=0.85cm, align=center, font=\footnotesize},
  arrow/.style={->, thick, >=stealth, shorten >= 1pt, shorten <= 1pt}
]
\node (T) at (3.0, 7.5) {$T$};
\node (D4) at (-1.8, 5.7) {$D_4$};
\node (D2d) at (0.2, 5.7) {$D_{2d}$};
\node (C4v) at (2.2, 5.7) {$C_{4v}$};
\node (D3) at (3.5, 4.25) {$D_3$};
\node (C3v) at (5.6, 4.25) {$C_{3v}$};
\node (C4) at (-1.8, 3.85) {$C_4$};
\node (D2) at (0.2, 3.85) {$D_2$};
\node (S4) at (1.2, 3.85) {$S_4$};
\node (C2v) at (2.2, 3.85) {$C_{2v}$};
\node (C3) at (4.6, 2.2) {$C_3$};
\node (C2) at (0.2, 1.8) {$C_2$};
\node (Cs) at (2.6, 1.8) {$C_s$};
\node (C1) at (2.3, 0.3) {$C_1$};

\draw[arrow] (C2) -- (C1); \draw[arrow] (Cs) -- (C1); \draw[arrow] (C3) -- (C1);
\draw[arrow] (C4) -- (C2); \draw[arrow] (D2) -- (C2); \draw[arrow] (S4) -- (C2);
\draw[arrow] (C2v) -- (C2); \draw[arrow] (D3) -- (C2);
\draw[arrow] (C2v) -- (Cs); \draw[arrow] (C3v) -- (Cs);
\draw[arrow] (D3) -- (C3); \draw[arrow] (C3v) -- (C3);
\draw[arrow] (T) to[bend left=16] (C3);
\draw[arrow] (D4) -- (C4); \draw[arrow] (C4v) -- (C4);
\draw[arrow] (D4) -- (D2); \draw[arrow] (D2d) -- (D2); \draw[arrow] (T) -- (D2);
\draw[arrow] (D2d) -- (S4);
\draw[arrow] (C4v) -- (C2v); \draw[arrow] (D2d) -- (C2v);
\end{tikzpicture}
\caption{Subgroup-type Hasse diagram of the 14 ME-complete parent spatial point groups (with subgroups identified by crystallographic point-group type and vertically ordered by group order). Arrows indicate subgroup inclusions within the ME-complete subset. The family forms a down-set on the subgroup lattice, proving that representation completeness is robustly inherited upon symmetry-lowering subgroup reduction.}
\label{fig:hasse}
\end{figure}

We prove that parent-level representation completeness is hereditary under point-group subgroup reduction:

\begin{theorem}[Down-Set Theorem]
\label{thm:downset}
If a parent crystallographic spatial point group $G$ is ME-complete ($\Supp(\Gamma_{\mathrm{ME}}^{[2]}) = \mathrm{Irr}(G)$), then every crystallographic point-group subgroup $H \le G$ is also ME-complete; hence the 14 ME-complete parent spatial point groups form a down-set on the subgroup lattice (Fig.~\ref{fig:hasse}).
\end{theorem}

The physical mechanism ensuring this heredity is that point-group reduction preserves vector and determinant mappings ($\Gamma_{\mathrm{pol}}(H) \cong \Gamma_{\mathrm{pol}}(G)\downarrow_H$ and $\Gamma_{\mathrm{ps}}(H) \cong \Gamma_{\mathrm{ps}}(G)\downarrow_H$), ensuring that response spaces restrict faithfully. Combined with Frobenius reciprocity (\hyperref[app:proof]{Appendix~A}; see \hyperref[app:category_adjunction]{Appendix~C} for the corresponding restriction--induction relation), every irreducible representation of the subgroup remains represented in the restricted response space. Thus, ME-complete parent groups provide symmetry environments with no representation-level dark channels under any symmetry-lowering structural descent.

Equivalently, the 14/18 partition is governed by five minimal ME-incomplete obstructions---$C_i$ ($\bar{1}$), $C_6$ ($6$), $C_{3h}$ ($\bar{6}$), $O$ ($432$), and $T_d$ ($\bar{4}3m$)---each of which is ME-incomplete while all of its proper subgroups are ME-complete; a parent point group $G$ is ME-complete if and only if it contains none of these five groups as a subgroup. Physically, these five obstructions follow directly from the inversion-odd spherical-tensor decomposition $V_{\mathrm{ME}} \cong (D_0^- \oplus D_1^- \oplus D_2^-)\downarrow_G$: spatial inversion ($C_i$) excludes all gerade modes across the 11 centrosymmetric groups, while the four noncentrosymmetric minimal groups host one-dimensional dark irreps whose first occurrence in this odd-parity series is at angular rank $J=3$---specifically the $|m|=3$ sectors of $C_6$ ($B$) and $C_{3h}$ ($A'$) and the cubic $xyz$ sector of $O$ ($A_2$) and $T_d$ ($A_1$). Here the spherical-tensor rank $J$ is distinct from the expansion order in the magnetic order parameter.

This global parent-group property bridges directly to downstream Shubnikov magnetic groups:

\begin{corollary}[Universal Non-Gray ME Activation for $\mathbf{k}=0$]
\label{cor:nongrey}
Let $G$ be an ME-complete parent spatial point group. For any commensurate $\mathbf{k}=0$ magnetic point group $M \le \mathcal{G}_{\mathrm{P}}$, linear magnetoelectric response is symmetry-allowed ($\operatorname{Fix}_M(V_{\mathrm{ME}}) \neq \{0\}$) if and only if $M$ is non-gray ($M \in \{\text{Type~I}, \text{Type~III}\}$).
\end{corollary}

\begin{proof}
Let $K = \pi_{\mathrm{spatial}}(M) \le G$. By Theorem~\ref{thm:downset}, $K$ inherits ME completeness ($\Supp(\Gamma_{\mathrm{ME}}^{[2]}(K)) = \mathrm{Irr}(K)$). For Type~I ($M=K$), the invariant tensor subspace satisfies $\operatorname{Fix}_M(V_{\mathrm{ME}}) \cong \operatorname{Hom}_K(\mathbf{1}_K, V_{\mathrm{ME}}) \neq \{0\}$ because the trivial irrep $\mathbf{1}_K$ is contained in the complete support $\Supp(\Gamma_{\mathrm{ME}}^{[2]}(K)) = \mathrm{Irr}(K)$. For Type~III ($M = H + \mathcal{T}(K-H)$ with $[K:H]=2$), invariance under $(K-H)\mathcal{T}$ requires the spatial operation to transform as the 1D alternating character $\chi_M: K \to \{\pm 1\}$; completeness of $K$ ensures $\operatorname{Fix}_M(V_{\mathrm{ME}}) \cong \operatorname{Hom}_K(\chi_M, V_{\mathrm{ME}}) \neq \{0\}$. Conversely, if $M$ is Type~II (gray), pure time-reversal $\mathcal{T} \in M$ enforces $\operatorname{Fix}_M(V_{\mathrm{ME}}) = \{0\}$ identically because $\tau_{\mathcal{T}}(V_{\mathrm{ME}}) = -1$.
\end{proof}

We emphasize that this 14/18 partition is a practical symmetry-screening tool rather than a ranking of magnetoelectric coupling strength: in the 18 ME-incomplete parents, linear ME activity still occurs whenever magnetic order condenses within the active partial support $\mathcal{S}_{\mathrm{act}}^{\mathrm{mat}}$ (as exemplified by $A_{1u}$ in incomplete $D_{3d}$ for Cr$_2$O$_3$ \cite{Rivera1994,Schmid1994,Litvin2014}). For first-principles and materials-discovery workflows, the tables in Appendix~\ref{app:tables} directly filter candidate magnetic configurations, minimize the number of independent response coefficients to compute, and provide consistency checks against assumed magnetic structures.

\textit{Three Meanings of Magnetoelectric Completeness.---}
To prevent confusion, three distinct questions enter this framework:
\begin{itemize}
\item \emph{Parent completeness}: Does $\Gamma_{\mathrm{ME}}^{[2]}$ contain every magnetic irrep of $G$? This holds for 14 of the 32 point groups, independently of the occupied magnetic sites.
\item \emph{Completeness of the constitutive basis}: For a selected sublattice-available active mode, have all independent $O(L)$ tensor templates been constructed? Spanning $\operatorname{Hom}_G(\Gamma_L, V_{\mathrm{ME}})$ answers this question and fixes the coefficient count $N(\Gamma_L)$.
\item \emph{Completeness at a selected OPD}: Do those evaluated templates exhaust the residual MPG tensor space? This holds for every 1D real sign irrep. For multidimensional modes it depends on the branch: the $\mathrm{Cr_2O_3}$ spin-flop and $\mathrm{BiFeO_3}$ in-plane examples at $\phi=\pi/2$ give 3D subspaces within 5D spaces, whereas the $T$ mode along $[001]$ in $\mathrm{Cu_2OSeO_3}$ spans the full 2D MPG space.
\end{itemize}
These distinguish representation support, a complete tensor basis, and its evaluated span; sublattice availability remains the separate material-specific screening condition.

\textit{Scope Boundary: $\mathbf{k}=0$ vs Finite-$\mathbf{k}$ and Spin-Point-Group Limit.---} The present construction is restricted to homogeneous $\mathbf{k}=0$ magnetic sectors within relativistic magnetic crystal symmetry (Shubnikov groups). Modulated finite-$\mathbf{k}$ order requires translation-sensitive little-group representations and, for nonsymmorphic systems, projective or corepresentational extensions; conversely, in the nonrelativistic zero-spin-orbit-coupling limit (or weakly spin-orbit-coupled regimes where independent spin rotations act as approximate symmetries), spin-point-group constraints can further restrict the allowed ME tensor \cite{Etxebarria2025}.

\subsection{\texorpdfstring{Beyond the $\mathcal{I}'$ Condition: Representation-Level Dark Channels}{Beyond the I' Condition: Representation-Level Dark Channels}}

At the present leading-order homogeneous level, the representation support criterion defines symmetry-enforced ``dark channels'' as parent magnetic irreps absent from $\Supp(\Gamma_{\mathrm{ME}}^{[2]})$, for which $\operatorname{Hom}_G(\Gamma_L, V_{\mathrm{ME}}) = \{0\}$. In hexagonal families ($C_6, D_6, C_{6v}$ in Table~\ref{tab:me_incomplete_active}), $B$-type irreps are missing from $\Supp(\Gamma_{\mathrm{ME}}^{[2]})$; any condensed $B$-mode generates an MPG containing $6'$, which constitutes a symmetry-enforced dark channel for linear ME response, consistent with the $C_{6v}$ examples tabulated in Ref.~\cite{Suzuki2019}.

More profoundly, this criterion provides a transparent parent-representation origin for the insufficiency of the anti-inversion condition $\mathcal{I}' = \mathcal{T}\mathcal{I}$. In centrosymmetric crystals, while spatial inversion $\mathcal{I}$ forbids linear magnetoelectricity, the combined time-reversal-inversion operation $\mathcal{I}' = \mathcal{T}\mathcal{I}$ is compatible with a linear ME tensor \cite{LandauEDC} and is necessary for phases condensed from a single odd-parity ($u$) parent magnetic mode. However, as recognized in full magnetic-group and multipolar analyses \cite{WatanabeYanase2018,Yatsushiro2021}, the presence of $\mathcal{I}'$ alone does not suffice to guarantee a nonvanishing linear ME effect. Global evaluation of $\Supp(\Gamma_{\mathrm{ME}}^{[2]})$ identifies three representative centrosymmetric parent families hosting $\mathcal{I}'$-preserving yet ME-inactive magnetic point groups: $6'/m$ ($C_{6h}(C_{3h})$), $6'/mmm'$ ($D_{6h}(D_{3h})$, along with its complementary setting $6'/mm'm$), and $m'\bar{3}'m$ ($O_h(T_d)$), cataloged in Table~\ref{tab:me_inactive_with_irreps}. Although each contains $\mathcal{I}'$, all satisfy $\operatorname{Fix}_M(V_{\mathrm{ME}}) = \{0\}$. In our framework, this suppression is traced directly upstream: their generating parent modes ($B_u$, $B_{1u,2u}$, and $A_{2u}$, respectively) are representation-level dark channels missing from $\Supp(\Gamma_{\mathrm{ME}}^{[2]})$, corresponding to the zero Hom-space sector ($\operatorname{Hom}_G(\Gamma_L, V_{\mathrm{ME}}) = \{0\}$) enforced by higher-order rotational and reflection symmetries. In terms of the two obstruction mechanisms identified in Sec.~IV\,A, these $\mathcal{I}'$-preserving counterexamples arise because their centrosymmetric parents ($C_{6h}, D_{6h}, O_h$) combine spatial inversion ($C_i$) with one of the $J=3$ minimal obstructions ($C_6, C_{3h}, O, T_d$): although the odd-parity ($u$-type) modes bypass the $C_i$ parity filter, their lowest occurrence in the inversion-odd spherical-tensor series lies at $J=3$ ($|m|=3$ or cubic $xyz$), outside the $J \le 2$ linear ME response space.

Inspection of the complete 32-group catalog (Tables~\ref{tab:me_incomplete_active} and \ref{tab:me_inactive_with_irreps}) reveals a sharper physical distinction between whole-mode prohibition and leading-order component constraints. Every dark irrep in the 32 crystallographic point groups falls into one of two categories: either an inversion-even ($g$-type) mode in a centrosymmetric parent (which preserves $\mathcal{I}$ and thus forbids $\alpha_{ij}$ identically at all orders), or a one-dimensional real sign representation $\chi_D$ (including dark trivial irreps such as $A'$ in $C_{3h}$, $A_1'$ in $D_{3h}$, and $A_1$ in $T_d$) whose characters take values in $\{\pm 1\}$ and therefore satisfy $\chi_D^{2r+1} = \chi_D$ for every non-negative integer $r$. Because macroscopic time-reversal odd parity forbids even powers $O(L^{2r})$, no higher odd power $O(L^{2r+1})$ of a single dark mode can enter $V_{\mathrm{ME}}$ in a pure single-mode expansion without an additional symmetry-breaking bias or independently condensed mode. Combining this exhaustiveness with the injectivity of the evaluation map $\operatorname{ev}_{\mathbf{L}^{(0)}}$ yields an exact equivalence for any nonzero single irreducible magnetic mode $\mathbf{L}$ across all 32 point groups:
\begin{equation}
N(\Gamma_L) > 0 \iff \operatorname{Fix}_{M(\mathbf{L})}(V_{\mathrm{ME}}) \neq \{0\}.
\label{eq:single_mode_equivalence}
\end{equation}
Thus, for a specified pure single-mode phase, parent-group screening and residual MPG analysis agree on whether net bulk linear ME response is globally forbidden; no single-mode phase is symmetry-forbidden at $O(L)$ yet symmetry-allowed at higher order. By contrast, vanishing leading-order components of an active multidimensional mode (such as $\alpha_{zz}^{(1)}=0$ and $\alpha_{xx}^{(1)}+\alpha_{yy}^{(1)}=0$ in $\mathrm{BiFeO_3}$) are not forbidden at all orders and can be lifted at $O(L^3)$ by cubic combinations of the same order parameter [Eq.~\eqref{eq:BFO_cubic}].

\section{Conclusions}

We have formulated a two-tier parent-group framework for the net bulk, homogeneous, static linear magnetoelectric response to first order in a single irreducible $\mathbf{k}=0$ magnetic mode across all 32 crystallographic point groups. Combining the universal parent response channels with the magnetic Wyckoff sublattice modes systematically separates mode availability from constitutive symmetry compatibility, reduces channel counting to polar tensor products via pseudoscalar cancellation, and constructs explicit leading-order tensor templates resolved into monopolar, toroidal, and quadrupolar sectors that obey Frobenius norm and angle preservation under order-parameter rotation. Retaining the parent-group mode identity and expansion order identifies the tensor subspace generated at leading order and the additional directions permitted by the residual MPG, thereby distinguishing leading-order constitutive constraints from the full symmetry-allowed response.

Applied through this standardized workflow to representative materials, the framework yields concrete physical and experimental criteria: (i) comparing isostructural $\mathrm{Cr_2O_3}$ and $\alpha\text{-}\mathrm{Fe_2O_3}$ shows that the inversion parity of the spin arrangement---not weak ferromagnetism---controls net bulk linear ME activity, while multi-channel coupling means that a vanishing single component may reflect channel cancellation; the same framework identifies the purely antisymmetric ME fingerprint of a candidate $A_{2u}$ ($++--$) ordering on the corundum $4c$ sublattice; (ii) in $\mathrm{BiFeO_3}$ and $\mathrm{Cr_2O_3}$, the mode symmetry locks whether field-induced polarizations are transverse or collinear with the probe field and N\'{e}el vector, enforcing leading-order sum rules ($\alpha_{zz}^{(1)}=0, \, \alpha_{xx}^{(1)}+\alpha_{yy}^{(1)}=0$) across all in-plane orientations while identifying the symmetry-allowed cubic $O(L^3)\sin 3\phi$ channel that can complete the residual $m'$ space; and (iii) in chiral $\mathrm{Cu_2OSeO_3}$, three-dimensional orientation analysis shows that the entire leading-order family across orientations permitting $2, 3, 4,$ and $9$ residual-MPG parameters is governed by the same two parent couplings $(\lambda_S, \lambda_A)$, and that transverse polarization in a single geometry can arise purely from symmetric quadrupolar coupling, making $(\alpha_{ij}\pm\alpha_{ji})/2$ comparisons and magnetic pinning conditions essential for isolating toroidal response. Globally, the 14/18 parent-completeness classification, its five minimal obstructions ($C_i, C_6, C_{3h}, O, T_d$), and the single-mode equivalence between leading-order and residual-MPG whole-mode symmetry prohibition provide a symmetry-first, algorithm-ready frontend for screening candidate magnetic configurations and streamlining first-principles calculations \cite{Xu2024,Iniguez2008,WojdelIniguez2009}, laying the groundwork for extending parent-group constitutive generation to third-rank piezomagnetism (Part~III \cite{Liu_piezo}).

\begin{acknowledgments}
This work was supported by the Advanced Materials-National Science and Technology Major Project of China (Grant No. 2025ZD0620000) and the National Natural Science Foundation of China (Grant Nos. 62174136 and 62674192). H. B. Niu acknowledges financial support from the Youth Innovation Team of Shaanxi Universities.
\end{acknowledgments}

\appendix

\section{Representation-Theoretic Proofs: Real Channel Multiplicities, the Down-Set Theorem, and Two-Tier Screening Heredity}
\label{app:proof}
\setcounter{equation}{0}
\renewcommand{\theequation}{A\arabic{equation}}
\renewcommand{\theHequation}{A\arabic{equation}}
\setcounter{lemma}{0}
\renewcommand{\thelemma}{A\arabic{lemma}}
\renewcommand{\theHlemma}{A\arabic{lemma}}

\textit{Real Constitutive Channel Dimension via Schur's Lemma.---}
Because $\Gamma_{\mathrm{pol}}$ and $\Gamma_{\mathrm{axial}}$ are real orthogonal representations on $\mathbb{R}^3$, the nine-dimensional response space $V_{\mathrm{ME}}^{\mathbb{R}}$ is a real orthogonal representation of $G$, and its complexification $V_{\mathrm{ME}} = V_{\mathrm{ME}}^{\mathbb{R}} \otimes_{\mathbb{R}} \mathbb{C}$ has a real-valued character $\chi_{\mathrm{ME}}(g) \in \mathbb{R}$ ($V_{\mathrm{ME}}^* \cong V_{\mathrm{ME}}$). For any physically irreducible real representation $U \in \mathrm{Irr}_{\mathbb{R}}(G)$ appearing with multiplicity $m_U$ in $V_{\mathrm{ME}}^{\mathbb{R}}$, Schur's lemma over $\mathbb{R}$ dictates that its endomorphism algebra $\operatorname{End}_G^{\mathbb{R}}(U)$ is isomorphic to $\mathbb{R}$ (real-type irrep) or $\mathbb{C}$ (complex-type real doublet, $U \otimes_{\mathbb{R}} \mathbb{C} \cong \sigma \oplus \sigma^*$ with $\sigma \not\cong \sigma^*$ in $\mathrm{Irr}_{\mathbb{C}}(G)$; quaternionic-type single-valued irreps do not occur in crystallographic point groups). The real dimension of the equivariant constitutive mapping space is therefore:
\begin{equation}
\dim_{\mathbb{R}}\operatorname{Hom}_G^{\mathbb{R}}\left(U, V_{\mathrm{ME}}^{\mathbb{R}}\right) = m_U \dim_{\mathbb{R}}\operatorname{End}_G^{\mathbb{R}}(U).
\label{eq:appA_real_Schur}
\end{equation}
For any real-type irrep ($\dim_{\mathbb{R}}\operatorname{End}_G^{\mathbb{R}}(U) = 1$, which encompasses all condensed order-parameter sectors explicitly analyzed in our four benchmark materials, whereas the uncondensed $E$ sector of point group $T$ is a complex-type real doublet), $m_U = \langle \chi_U, \chi_{\mathrm{ME}} \rangle_G = N$, so $N$ directly counts the independent real Cartesian tensor templates. For a complex-type 2D real doublet $U$ (such as the $E$ doublets in cyclic or chiral groups $C_3, C_4, S_4, C_6, C_{3h}, C_{4h}, C_{6h}, C_{3i}, T, T_h$), $\operatorname{End}_G^{\mathbb{R}}(U) \cong \mathbb{C}$ is two-dimensional over $\mathbb{R}$ (spanned by the identity $I_2$ and the $90^\circ$ rotation $J = \begin{pmatrix} 0 & -1 \\ 1 & 0 \end{pmatrix}$), while reality of $\chi_{\mathrm{ME}}$ ensures $n_\sigma^{\mathrm{ME}} = n_{\sigma^*}^{\mathrm{ME}} = m_U$; thus each doublet copy in $V_{\mathrm{ME}}^{\mathbb{R}}$ supports two real couplings ($a\mathbf{L} + bJ\mathbf{L}$), yielding $\dim_{\mathbb{R}}\operatorname{Hom}_G^{\mathbb{R}}(U, V_{\mathrm{ME}}^{\mathbb{R}}) = 2m_U = n_\sigma^{\mathrm{ME}} + n_{\sigma^*}^{\mathrm{ME}}$ independent real parameters.

\textit{Proof of the Down-Set Theorem.---}
Let $\mathrm{Irr}(G)$ denote the set of inequivalent irreducible representations of a finite group $G$. For any representation $\Gamma$, its support is $\Supp(\Gamma) \equiv \{ \rho \in \mathrm{Irr}(G) \mid \langle \rho, \Gamma \rangle > 0 \}$.

\begin{lemma}[Support Heredity]
\label{lem:support_heredity}
Let $\Gamma$ be a representation of $G$. If $\Supp(\Gamma) = \mathrm{Irr}(G)$, then for any subgroup $H \le G$, the restriction satisfies $\Supp(\Gamma \downarrow_H) = \mathrm{Irr}(H)$.
\end{lemma}

\begin{proof}
Let \mbox{$\sigma \in \mathrm{Irr}(H)$} be an arbitrary irrep of $H$. The induced representation $\operatorname{Ind}_H^G \sigma$ decomposes into irreps of $G$. Thus, there exists at least one $\rho \in \mathrm{Irr}(G)$ such that $\langle \rho, \operatorname{Ind}_H^G \sigma \rangle_G > 0$. By Frobenius reciprocity, $\langle \rho \downarrow_H, \sigma \rangle_H > 0$, implying $\sigma \subset \rho \downarrow_H$. Since $\Supp(\Gamma) = \mathrm{Irr}(G)$ by assumption, $\rho \subset \Gamma$. Consequently, $\sigma \subset \rho \downarrow_H \subset \Gamma \downarrow_H$. Because this holds for every $\sigma \in \mathrm{Irr}(H)$, $\Supp(\Gamma \downarrow_H) = \mathrm{Irr}(H)$.
\end{proof}

\begin{proof}[Proof of the Down-Set Theorem (Theorem~\ref{thm:downset})]
The rank-two linear ME representation is $\Gamma_{\mathrm{ME}}^{[2]}(G) = \Gamma_{\mathrm{pol}}(G) \otimes \Gamma_{\mathrm{pol}}(G) \otimes \Gamma_{\mathrm{ps}}(G)$. Upon symmetry reduction to $H \le G$, the vector space and determinant mapping enforce $\Gamma_{\mathrm{pol}}(H) \cong \Gamma_{\mathrm{pol}}(G) \downarrow_H$ and $\Gamma_{\mathrm{ps}}(H) \cong \Gamma_{\mathrm{ps}}(G) \downarrow_H$. Therefore, $\Gamma_{\mathrm{ME}}^{[2]}(H) \cong \Gamma_{\mathrm{ME}}^{[2]}(G) \downarrow_H$. If $G$ is ME-complete, applying Lemma~\ref{lem:support_heredity} immediately yields $\Supp(\Gamma_{\mathrm{ME}}^{[2]}(H)) = \mathrm{Irr}(H)$. Thus, $H$ is also ME-complete.
\end{proof}

\textit{Consequence for Two-Tier Sublattice Screening.---}
In a specific crystal with $G$-invariant magnetic Wyckoff sublattice $\mathcal{S}_{\mathrm{mag}}$ held fixed and sublattice magnetic representation $\Gamma_{\mathrm{mag}}^{\mathrm{mat}}$ [Eq.~\eqref{eq:sublattice_mag_rep}], the material-active sector set under any point-group subgroup $H \le G$ (obtained by restricting the same $G$-equivariant sublattice representation to $H$, whereas a reconstruction of the magnetic Wyckoff orbit across a structural phase transition requires a new Tier-2 representation) is governed by the two-tier intersection $\mathcal{S}_{\mathrm{act}}^{\mathrm{mat}}(H) = \Supp(\Gamma_{\mathrm{mag}}^{\mathrm{mat}}\downarrow_H) \cap \Supp(\Gamma_{\mathrm{ME}}^{[2]}(H))$. When the parent group $G$ is ME-complete, Theorem~\ref{thm:downset} guarantees $\Supp(\Gamma_{\mathrm{ME}}^{[2]}(H)) = \mathrm{Irr}(H)$ for every $H \le G$, so the two-tier intersection reduces identically to the sublattice kinematic support:
\begin{equation}
\mathcal{S}_{\mathrm{act}}^{\mathrm{mat}}(H) = \Supp\left(\Gamma_{\mathrm{mag}}^{\mathrm{mat}}\downarrow_H\right) \cap \mathrm{Irr}(H) = \Supp\left(\Gamma_{\mathrm{mag}}^{\mathrm{mat}}\downarrow_H\right).
\label{eq:appA_twotier}
\end{equation}
Consequently, Class~B (sublattice-available but linear-ME-dark modes, $\Supp(\Gamma_{\mathrm{mag}}^{\mathrm{mat}}\downarrow_H) \setminus \Supp(\Gamma_{\mathrm{ME}}^{[2]}(H))$) is identically empty across the entire subgroup lattice of an ME-complete parent group. Conversely, for the 18 ME-incomplete parent groups where $\mathcal{S}_{\mathrm{ME}}(G) \subsetneq \mathrm{Irr}(G)$, parent symmetry acts as a potential filter on $\Supp(\Gamma_{\mathrm{mag}}^{\mathrm{mat}})$, producing a non-empty Class~B whenever $\Supp(\Gamma_{\mathrm{mag}}^{\mathrm{mat}}) \cap \mathcal{S}_{\mathrm{dark}}(G) \neq \varnothing$.

\section{Magnetoelectric Completeness and Two-Tier Support Screening across the 32 Crystallographic Point Groups}
\label{app:tables}
\setcounter{table}{0}
\renewcommand{\thetable}{B\arabic{table}}
\renewcommand{\theHtable}{B\arabic{table}}

\begin{table*}[!p]
\caption{\textbf{Tier-1 ME-Complete Parent Point Groups (14 Groups, \cR{red}):} Parent point groups satisfying $\mathcal{S}_{\mathrm{ME}}(G) = \mathrm{Irr}(G)$ and $\mathcal{S}_{\mathrm{dark}}(G) = \varnothing$. Every sublattice-available mode belongs to Class~A ($\mathcal{S}_{\mathrm{act}}^{\mathrm{mat}} = \mathcal{S}_{\mathrm{mag}}^{\mathrm{mat}}$). Column~3 lists the ME-active 1D-derived non-gray MPGs (\cB{blue}) in unified $G(H)$ notation ($G(G)$ for Type-I and $G(H)$ with $[G:H]=2$ for Type-III).}
\label{tab:me_complete_induced}
\centering
\footnotesize
\setlength{\tabcolsep}{4pt}
\renewcommand{\arraystretch}{0.92}
\begin{tabular*}{\textwidth}{@{\extracolsep{\fill}}l p{8.3cm} p{6.4cm}@{}}
\toprule
\textbf{Parent $G$} & \RaggedCell{\textbf{Tier-1 Active Multipoles $\Gamma_{\mathrm{ME}}^{[2]} = [\theta] \oplus [\mathbf{T}_\alpha] \oplus [\mathbf{Q}_S]$ ($\mathcal{S}_{\mathrm{dark}} = \varnothing$)}} & \RaggedCell{\textbf{Active 1D-Derived MPGs \mbox{$G(H)$ [$\Gamma_L$]}}} \\
\midrule
\cR{$C_1 \ (1)$}            & \RaggedCell{$A[\theta] \oplus 3A[\mathbf{T}_\alpha] \oplus 5A[\mathbf{Q}_S]$} & \RaggedCell{\mbox{\cB{$C_1(C_1)$} [$A$]}} \\
\cR{$C_2 \ (2)$}            & \RaggedCell{$A[\theta] \oplus (A\oplus 2B)[\mathbf{T}_\alpha] \oplus (3A\oplus 2B)[\mathbf{Q}_S]$} & \RaggedCell{\mbox{\cB{$C_2(C_2)$} [$A$]}; \mbox{\cB{$C_2(C_1)$} [$B$]}} \\
\cR{$C_s \ (m)$}            & \RaggedCell{$A''[\theta] \oplus (2A'\oplus A'')[\mathbf{T}_\alpha] \oplus (2A'\oplus 3A'')[\mathbf{Q}_S]$} & \RaggedCell{\mbox{\cB{$C_s(C_s)$} [$A'$]}; \mbox{\cB{$C_s(C_1)$} [$A''$]}} \\
\cR{$D_2 \ (222)$}          & \RaggedCell{$A[\theta] \oplus (B_1\oplus B_2\oplus B_3)[\mathbf{T}_\alpha] \oplus (2A\oplus B_1\oplus B_2\oplus B_3)[\mathbf{Q}_S]$} & \RaggedCell{\mbox{\cB{$D_2(D_2)$} [$A$]}; \mbox{\cB{$D_2(C_2)$} [$B_{1,2,3}$]}} \\
\cR{$C_{2v} \ (mm2)$}       & \RaggedCell{$A_2[\theta] \oplus (A_1\oplus B_1\oplus B_2)[\mathbf{T}_\alpha] \oplus (A_1\oplus 2A_2\oplus B_1\oplus B_2)[\mathbf{Q}_S]$} & \RaggedCell{\mbox{\cB{$C_{2v}(C_{2v})$} [$A_1$]}; \mbox{\cB{$C_{2v}(C_2)$} [$A_2$]}; \mbox{\cB{$C_{2v}(C_s)$} [$B_{1,2}$]}} \\
\cR{$C_3 \ (3)$}            & \RaggedCell{$A[\theta] \oplus (A\oplus E)[\mathbf{T}_\alpha] \oplus (A\oplus 2E)[\mathbf{Q}_S]$} & \RaggedCell{\mbox{\cB{$C_3(C_3)$} [$A$]}} \\
\cR{$D_3 \ (32)$}           & \RaggedCell{$A_1[\theta] \oplus (A_2\oplus E)[\mathbf{T}_\alpha] \oplus (A_1\oplus 2E)[\mathbf{Q}_S]$} & \RaggedCell{\mbox{\cB{$D_3(D_3)$} [$A_1$]}; \mbox{\cB{$D_3(C_3)$} [$A_2$]}} \\
\cR{$C_{3v} \ (3m)$}        & \RaggedCell{$A_2[\theta] \oplus (A_1\oplus E)[\mathbf{T}_\alpha] \oplus (A_2\oplus 2E)[\mathbf{Q}_S]$} & \RaggedCell{\mbox{\cB{$C_{3v}(C_{3v})$} [$A_1$]}; \mbox{\cB{$C_{3v}(C_3)$} [$A_2$]}} \\
\cR{$C_4 \ (4)$}            & \RaggedCell{$A[\theta] \oplus (A\oplus E)[\mathbf{T}_\alpha] \oplus (A\oplus 2B\oplus E)[\mathbf{Q}_S]$} & \RaggedCell{\mbox{\cB{$C_4(C_4)$} [$A$]}; \mbox{\cB{$C_4(C_2)$} [$B$]}} \\
\cR{$S_4 \ (\bar{4})$}      & \RaggedCell{$B[\theta] \oplus (B\oplus E)[\mathbf{T}_\alpha] \oplus (2A\oplus B\oplus E)[\mathbf{Q}_S]$} & \RaggedCell{\mbox{\cB{$S_4(S_4)$} [$A$]}; \mbox{\cB{$S_4(C_2)$} [$B$]}} \\
\cR{$D_4 \ (422)$}          & \RaggedCell{$A_1[\theta] \oplus (A_2\oplus E)[\mathbf{T}_\alpha] \oplus (A_1\oplus B_1\oplus B_2\oplus E)[\mathbf{Q}_S]$} & \RaggedCell{\mbox{\cB{$D_4(D_4)$} [$A_1$]}; \mbox{\cB{$D_4(C_4)$} [$A_2$]}; \mbox{\cB{$D_4(D_2)$} [$B_{1,2}$]}} \\
\cR{$C_{4v} \ (4mm)$}       & \RaggedCell{$A_2[\theta] \oplus (A_1\oplus E)[\mathbf{T}_\alpha] \oplus (A_2\oplus B_1\oplus B_2\oplus E)[\mathbf{Q}_S]$} & \RaggedCell{\mbox{\cB{$C_{4v}(C_{4v})$} [$A_1$]}; \mbox{\cB{$C_{4v}(C_4)$} [$A_2$]}; \mbox{\cB{$C_{4v}(C_{2v})$} [$B_{1,2}$]}} \\
\cR{$D_{2d} \ (\bar{4}2m)$} & \RaggedCell{$B_1[\theta] \oplus (B_2\oplus E)[\mathbf{T}_\alpha] \oplus (A_1\oplus A_2\oplus B_1\oplus E)[\mathbf{Q}_S]$} & \RaggedCell{\mbox{\cB{$D_{2d}(D_{2d})$} [$A_1$]}; \mbox{\cB{$D_{2d}(S_4)$} [$A_2$]}; \mbox{\cB{$D_{2d}(D_2)$} [$B_1$]}; \mbox{\cB{$D_{2d}(C_{2v})$} [$B_2$]}} \\
\cR{$T \ (23)$}             & \RaggedCell{$A[\theta] \oplus T[\mathbf{T}_\alpha] \oplus (E\oplus T)[\mathbf{Q}_S]$} & \RaggedCell{\mbox{\cB{$T(T)$} [$A$]}} \\
\bottomrule
\end{tabular*}

\vspace{1.4em}

\caption{\textbf{Non-Centrosymmetric ME-Incomplete Parent Point Groups (7 Groups, \cO{orange}):} Non-centrosymmetric parents with $\mathcal{S}_{\mathrm{ME}}(G) \subsetneq \mathrm{Irr}(G)$. All 2D/3D irreps are active in Column~2; Column~3 lists ME-active 1D-derived MPGs (\cB{blue}); Column~4 lists the unsupported 1D dark irreps $\mathcal{S}_{\mathrm{dark}}(G)$ and their induced dark MPGs $G(H)$.}
\label{tab:me_incomplete_active}
\centering
\footnotesize
\setlength{\tabcolsep}{3.5pt}
\renewcommand{\arraystretch}{0.92}
\begin{tabular*}{\textwidth}{@{\extracolsep{\fill}}l p{6.5cm} p{3.6cm} p{4.8cm}@{}}
\toprule
\textbf{Parent $G$} & \RaggedCell{\textbf{Tier-1 Active Multipoles $\Gamma_{\mathrm{ME}}^{[2]} = [\theta] \oplus [\mathbf{T}_\alpha] \oplus [\mathbf{Q}_S]$}} & \RaggedCell{\textbf{Active 1D MPGs \mbox{$G(H)$ [$\Gamma_L$]}}} & \RaggedCell{\textbf{Tier-1 Dark Irreps $\mathcal{S}_{\mathrm{dark}}(G)$ [Dark 1D MPG $G(H)$]}} \\
\midrule
\cO{$C_6 \ (6)$}            & \RaggedCell{$A[\theta] \oplus (A\oplus E_1)[\mathbf{T}_\alpha] \oplus (A\oplus E_1\oplus E_2)[\mathbf{Q}_S]$} & \RaggedCell{\mbox{\cB{$C_6(C_6)$} [$A$]}} & \RaggedCell{\mbox{$B$ [$C_6(C_3) = 6'$]}} \\
\cO{$D_6 \ (622)$}          & \RaggedCell{$A_1[\theta] \oplus (A_2\oplus E_1)[\mathbf{T}_\alpha] \oplus (A_1\oplus E_1\oplus E_2)[\mathbf{Q}_S]$} & \RaggedCell{\mbox{\cB{$D_6(D_6)$} [$A_1$]}; \mbox{\cB{$D_6(C_6)$} [$A_2$]}} & \RaggedCell{\mbox{$B_{1,2}$ [$D_6(D_3) = 6'22', 6'2'2$]}} \\
\cO{$C_{6v} \ (6mm)$}       & \RaggedCell{$A_2[\theta] \oplus (A_1\oplus E_1)[\mathbf{T}_\alpha] \oplus (A_2\oplus E_1\oplus E_2)[\mathbf{Q}_S]$} & \RaggedCell{\mbox{\cB{$C_{6v}(C_{6v})$} [$A_1$]}; \mbox{\cB{$C_{6v}(C_6)$} [$A_2$]}} & \RaggedCell{\mbox{$B_{1,2}$ [$C_{6v}(C_{3v}) = 6'mm', 6'm'm$]}} \\
\cO{$O \ (432)$}            & \RaggedCell{$A_1[\theta] \oplus T_1[\mathbf{T}_\alpha] \oplus (E\oplus T_2)[\mathbf{Q}_S]$} & \RaggedCell{\mbox{\cB{$O(O)$} [$A_1$]}} & \RaggedCell{\mbox{$A_2$ [$O(T) = 4'32'$]}} \\
\cO{$C_{3h} \ (\bar{6})$}   & \RaggedCell{$A''[\theta] \oplus (A''\oplus E')[\mathbf{T}_\alpha] \oplus (A''\oplus E'\oplus E'')[\mathbf{Q}_S]$} & \RaggedCell{\mbox{\cB{$C_{3h}(C_3)$} [$A''$]}} & \RaggedCell{\mbox{$A'$ [$C_{3h}(C_{3h}) = \bar{6}$]}} \\
\cO{$D_{3h} \ (\bar{6}m2)$} & \RaggedCell{$A_1''[\theta] \oplus (A_2''\oplus E')[\mathbf{T}_\alpha] \oplus (A_1''\oplus E'\oplus E'')[\mathbf{Q}_S]$} & \RaggedCell{\mbox{\cB{$D_{3h}(D_3)$} [$A_1''$]}; \mbox{\cB{$D_{3h}(C_{3v})$} [$A_2''$]}} & \RaggedCell{\mbox{$A_1'$ [$D_{3h}(D_{3h})$]}; \mbox{$A_2'$ [$D_{3h}(C_{3h})$]}} \\
\cO{$T_d \ (\bar{4}3m)$}    & \RaggedCell{$A_2[\theta] \oplus T_2[\mathbf{T}_\alpha] \oplus (E\oplus T_1)[\mathbf{Q}_S]$} & \RaggedCell{\mbox{\cB{$T_d(T)$} [$A_2$]}} & \RaggedCell{\mbox{$A_1$ [$T_d(T_d) = \bar{4}3m$]}} \\
\bottomrule
\end{tabular*}
\end{table*}

\begin{table*}[!t]
\caption{\textbf{Centrosymmetric ME-Incomplete Parent Point Groups (11 Groups, \cO{orange}):} Centrosymmetric parents with $\mathcal{S}_{\mathrm{ME}}(G) \subseteq \mathrm{Irr}_u(G) \subsetneq \mathrm{Irr}(G)$. Column~2 lists the active multipolar decomposition; Column~3 lists ME-active 1D-derived MPGs (\cB{blue}); Column~4 partitions $\mathcal{S}_{\mathrm{dark}}(G) = \mathcal{S}_{u\text{-dark}}(G) \cup \mathcal{S}_{g\text{-dark}}(G)$ into odd-parity \textbf{$u$-Dark} counterexamples (preserving $\mathcal{I}' = \mathcal{T}\mathcal{I}$ yet yielding dark MPGs $6'/m$, \mbox{$6'/mmm'$ [$6'/mm'm$]}, and $m'\bar{3}'m$) and even-parity \textbf{$g$-Dark} channels.}
\label{tab:me_inactive_with_irreps}
\centering
\footnotesize
\setlength{\tabcolsep}{3.5pt}
\renewcommand{\arraystretch}{0.96}
\begin{tabular*}{\textwidth}{@{\extracolsep{\fill}}l p{6.3cm} p{3.4cm} p{5.2cm}@{}}
\toprule
\textbf{Parent $G$} & \RaggedCell{\textbf{Tier-1 Active Multipoles $\Gamma_{\mathrm{ME}}^{[2]} = [\theta] \oplus [\mathbf{T}_\alpha] \oplus [\mathbf{Q}_S]$}} & \RaggedCell{\textbf{Active 1D MPGs \mbox{$G(H)$ [$\Gamma_L$]}}} & \RaggedCell{\textbf{Tier-1 Dark Irreps $\mathcal{S}_{\mathrm{dark}}(G)$ ($u$-Dark vs.\ $g$-Dark \& Dark 1D MPGs $G(H)$)}} \\
\midrule
\cO{$C_i \ (\bar{1})$}      & \RaggedCell{$A_u[\theta] \oplus 3A_u[\mathbf{T}_\alpha] \oplus 5A_u[\mathbf{Q}_S]$} & \RaggedCell{\mbox{\cB{$C_i(C_1)$} [$A_u$]}} & \RaggedCell{\textbf{$u$-Dark:} \textit{None}; \textbf{$g$-Dark:} \mbox{$A_g$ [$C_i(C_i)$]}} \\
\cO{$C_{2h} \ (2/m)$}       & \RaggedCell{$A_u[\theta] \oplus (A_u\oplus 2B_u)[\mathbf{T}_\alpha] \oplus (3A_u\oplus 2B_u)[\mathbf{Q}_S]$} & \RaggedCell{\mbox{\cB{$C_{2h}(C_2)$} [$A_u$]}; \mbox{\cB{$C_{2h}(C_s)$} [$B_u$]}} & \RaggedCell{\textbf{$u$-Dark:} \textit{None}; \textbf{$g$-Dark:} \mbox{$A_g$ [$C_{2h}(C_{2h})$]}, \mbox{$B_g$ [$C_{2h}(C_i)$]}} \\
\cO{$D_{2h} \ (mmm)$}       & \RaggedCell{$A_u[\theta] \oplus (B_{1u}\oplus B_{2u}\oplus B_{3u})[\mathbf{T}_\alpha] \oplus (2A_u\oplus B_{1u}\oplus B_{2u}\oplus B_{3u})[\mathbf{Q}_S]$} & \RaggedCell{\mbox{\cB{$D_{2h}(D_2)$} [$A_u$]}; \mbox{\cB{$D_{2h}(C_{2v})$} [$B_{1u,2u,3u}$]}} & \RaggedCell{\textbf{$u$-Dark:} \textit{None}; \textbf{$g$-Dark:} \mbox{$A_g$ [$D_{2h}(D_{2h})$]}, \mbox{$B_{1g,2g,3g}$ [$D_{2h}(C_{2h})$]}} \\
\cO{$C_{4h} \ (4/m)$}       & \RaggedCell{$A_u[\theta] \oplus (A_u\oplus E_u)[\mathbf{T}_\alpha] \oplus (A_u\oplus 2B_u\oplus E_u)[\mathbf{Q}_S]$} & \RaggedCell{\mbox{\cB{$C_{4h}(C_4)$} [$A_u$]}; \mbox{\cB{$C_{4h}(S_4)$} [$B_u$]}} & \RaggedCell{\textbf{$u$-Dark:} \textit{None}; \textbf{$g$-Dark:} \mbox{$A_g$ [$C_{4h}(C_{4h})$]}, \mbox{$B_g$ [$C_{4h}(C_{2h})$]}, $E_g$} \\
\cO{$D_{4h} \ (4/mmm)$}     & \RaggedCell{$A_{1u}[\theta] \oplus (A_{2u}\oplus E_u)[\mathbf{T}_\alpha] \oplus (A_{1u}\oplus B_{1u}\oplus B_{2u}\oplus E_u)[\mathbf{Q}_S]$} & \RaggedCell{\mbox{\cB{$D_{4h}(D_4)$} [$A_{1u}$]}; \mbox{\cB{$D_{4h}(C_{4v})$} [$A_{2u}$]}; \mbox{\cB{$D_{4h}(D_{2d})$} [$B_{1u,2u}$]}} & \RaggedCell{\textbf{$u$-Dark:} \textit{None}; \textbf{$g$-Dark:} \mbox{$A_{1g}$ [$D_{4h}(D_{4h})$]}, \mbox{$A_{2g}$ [$D_{4h}(C_{4h})$]}, \mbox{$B_{1g,2g}$ [$D_{4h}(D_{2h})$]}, $E_g$} \\
\cO{$C_{3i} \ (\bar{3})$}   & \RaggedCell{$A_u[\theta] \oplus (A_u\oplus E_u)[\mathbf{T}_\alpha] \oplus (A_u\oplus 2E_u)[\mathbf{Q}_S]$} & \RaggedCell{\mbox{\cB{$C_{3i}(C_3)$} [$A_u$]}} & \RaggedCell{\textbf{$u$-Dark:} \textit{None}; \textbf{$g$-Dark:} \mbox{$A_g$ [$C_{3i}(C_{3i})$]}, $E_g$} \\
\cO{$D_{3d} \ (\bar{3}m)$}  & \RaggedCell{$A_{1u}[\theta] \oplus (A_{2u}\oplus E_u)[\mathbf{T}_\alpha] \oplus (A_{1u}\oplus 2E_u)[\mathbf{Q}_S]$} & \RaggedCell{\mbox{\cB{$D_{3d}(D_3)$} [$A_{1u}$]}; \mbox{\cB{$D_{3d}(C_{3v})$} [$A_{2u}$]}} & \RaggedCell{\textbf{$u$-Dark:} \textit{None}; \textbf{$g$-Dark:} \mbox{$A_{1g}$ [$D_{3d}(D_{3d})$]}, \mbox{$A_{2g}$ [$D_{3d}(C_{3i})$]}, $E_g$} \\
\cO{$C_{6h} \ (6/m)$}       & \RaggedCell{$A_u[\theta] \oplus (A_u\oplus E_{1u})[\mathbf{T}_\alpha] \oplus (A_u\oplus E_{1u}\oplus E_{2u})[\mathbf{Q}_S]$} & \RaggedCell{\mbox{\cB{$C_{6h}(C_6)$} [$A_u$]}} & \RaggedCell{\textbf{$u$-Dark ($\mathcal{I}'$):} \mbox{$B_u$ [$C_{6h}(C_{3h}) = 6'/m$]}; \newline \textbf{$g$-Dark:} \mbox{$A_g$ [$C_{6h}(C_{6h})$]}, \mbox{$B_g$ [$C_{6h}(C_{3i})$]}, $E_{1g,2g}$} \\
\cO{$D_{6h} \ (6/mmm)$}     & \RaggedCell{$A_{1u}[\theta] \oplus (A_{2u}\oplus E_{1u})[\mathbf{T}_\alpha] \oplus (A_{1u}\oplus E_{1u}\oplus E_{2u})[\mathbf{Q}_S]$} & \RaggedCell{\mbox{\cB{$D_{6h}(D_6)$} [$A_{1u}$]}; \mbox{\cB{$D_{6h}(C_{6v})$} [$A_{2u}$]}} & \RaggedCell{\textbf{$u$-Dark ($\mathcal{I}'$):} \mbox{$B_{1u,2u}$ [$D_{6h}(D_{3h})$}\newline $= 6'/mmm', 6'/mm'm$]; \newline \textbf{$g$-Dark:} \mbox{$A_{1g}$ [$D_{6h}(D_{6h})$]}, \mbox{$A_{2g}$ [$D_{6h}(C_{6h})$]}, \mbox{$B_{1g,2g}$ [$D_{6h}(D_{3d})$]}, $E_{1g,2g}$} \\
\cO{$T_h \ (m\bar{3})$}     & \RaggedCell{$A_u[\theta] \oplus T_u[\mathbf{T}_\alpha] \oplus (E_u\oplus T_u)[\mathbf{Q}_S]$} & \RaggedCell{\mbox{\cB{$T_h(T)$} [$A_u$]}} & \RaggedCell{\textbf{$u$-Dark:} \textit{None}; \textbf{$g$-Dark:} \mbox{$A_g$ [$T_h(T_h)$]}, $E_g, T_g$} \\
\cO{$O_h \ (m\bar{3}m)$}    & \RaggedCell{$A_{1u}[\theta] \oplus T_{1u}[\mathbf{T}_\alpha] \oplus (E_u\oplus T_{2u})[\mathbf{Q}_S]$} & \RaggedCell{\mbox{\cB{$O_h(O)$} [$A_{1u}$]}} & \RaggedCell{\textbf{$u$-Dark ($\mathcal{I}'$):} \mbox{$A_{2u}$ [$O_h(T_d) = m'\bar{3}'m$]}; \newline \textbf{$g$-Dark:} \mbox{$A_{1g}$ [$O_h(O_h)$]}, \mbox{$A_{2g}$ [$O_h(T_h) = m\bar{3}m'$]}, $E_g, T_{1g,2g}$} \\
\bottomrule
\end{tabular*}
\end{table*}

This Appendix provides the exhaustive Tier-1 parent-group representation screening of $\Gamma_{\mathrm{ME}}^{[2]} = D_0^-\downarrow_G \oplus D_1^-\downarrow_G \oplus D_2^-\downarrow_G$ (monopole $\theta$, toroidal vector $\mathbf{T}_\alpha$, and quadrupole $\mathbf{Q}_S$), the active support $\mathcal{S}_{\mathrm{ME}}(G) \equiv \Supp(\Gamma_{\mathrm{ME}}^{[2]})$, and the complementary dark irrep set $\mathcal{S}_{\mathrm{dark}}(G) \equiv \mathrm{Irr}(G) \setminus \mathcal{S}_{\mathrm{ME}}(G)$ across all 32 crystallographic point groups, partitioning them into 14 ME-complete (\cR{red}) and 18 ME-incomplete (\cO{orange}) parent groups ($32 = 14 + 7 + 11$). Within the two-tier screening protocol [Eq.~\eqref{eq:two_tier_intersection}], Tables~\ref{tab:me_complete_induced}--\ref{tab:me_inactive_with_irreps} serve as universal Tier-1 lookup tables to be intersected with a given crystal's Tier-2 sublattice kinematic support $\mathcal{S}_{\mathrm{mag}}^{\mathrm{mat}} = \Supp(\Gamma_{\mathrm{mag}}^{\mathrm{mat}})$:
(i)~\emph{Tier-2 lookup protocol (Classes A, B, C, and D):} Intersecting $\mathcal{S}_{\mathrm{mag}}^{\mathrm{mat}}$ with $\mathcal{S}_{\mathrm{ME}}(G)$ in Tables~\ref{tab:me_complete_induced}--\ref{tab:me_inactive_with_irreps} partitions all irreps of $G$ into four exhaustive classes: \textbf{Class~A} ($\rho \in \mathcal{S}_{\mathrm{mag}}^{\mathrm{mat}} \cap \mathcal{S}_{\mathrm{ME}}(G)$: sublattice-available and ME-active), \textbf{Class~B} ($\rho \in \mathcal{S}_{\mathrm{mag}}^{\mathrm{mat}} \cap \mathcal{S}_{\mathrm{dark}}(G)$: sublattice-available but linear-ME dark), \textbf{Class~C} ($\rho \in \mathcal{S}_{\mathrm{ME}}(G) \setminus \mathcal{S}_{\mathrm{mag}}^{\mathrm{mat}}$: parent-allowed but unavailable on $\mathcal{S}_{\mathrm{mag}}$), and \textbf{Class~D} ($\rho \notin \mathcal{S}_{\mathrm{mag}}^{\mathrm{mat}} \cup \mathcal{S}_{\mathrm{ME}}(G)$: absent from both supports and excluded from the candidate space).
(ii)~\emph{Multipolar decomposition and unified $G(H)$ MPG notation:} Column~2 lists exclusively the active multipolar decomposition $\Gamma_{\mathrm{ME}}^{[2]} = [\theta] \oplus [\mathbf{T}_\alpha] \oplus [\mathbf{Q}_S]$ using standard physical irrep labels $\mathrm{Irr}_{\mathbb{R}}(G)$ (where conjugate 1D irreps $\sigma, \sigma^* \in \mathrm{Irr}_{\mathbb{C}}(G)$ in $C_3, C_4, S_4, C_6, C_{3h}, C_{4h}, C_{6h}, C_{3i}, T, T_h$ are denoted by their 2D real doublet symbol $E$, each copy supporting two real parameters per Appendix~\ref{app:proof}). Multidimensional irreps ($E, T$) are fully included in Column~2 (and in $\mathcal{S}_{\mathrm{dark}}(G)$ of Column~4); because their isotropy subgroups depend on the selected OPD, Column~3 (and brackets in Column~4) lists specifically the non-gray MPGs uniquely induced by 1D irreps $\Gamma_L$ using the unified kernel-subgroup notation $G(H)$ with $H = \ker\Gamma_L \le G$: $G(G) \equiv G$ denotes a Type-I MPG ($H=G$, $\Gamma_L = \Gamma_1$), whereas $G(H)$ with $[G:H]=2$ denotes a Type-III MPG (active 1D-derived MPGs highlighted in \cB{blue}).
(iii)~\emph{Non-centrosymmetric vs.\ ME-complete and $B_1, B_2$ settings:} Table~\ref{tab:me_incomplete_active} shows that the absence of inversion removes the universal $g/u$ parity obstruction (Table~\ref{tab:me_inactive_with_irreps}) but does not guarantee ME completeness, as rotational/reflection symmetries still produce 1D dark sectors. Here $H$ denotes an embedded kernel in the chosen crystal-axis setting; identical Schoenflies labels need not specify identical embeddings \cite{Kopsky2006,Litvin2014}. In $D_{2d}$ ($\bar{4}2m$), $B_1$ and $B_2$ yield distinct magnetic point-group types $D_{2d}(D_2) = \bar{4}'2m'$ and $D_{2d}(C_{2v}) = \bar{4}'2'm$; in $C_{2v}, D_4, C_{4v}, D_{4h}, D_6, C_{6v}, D_{6h}$ (and $B_{1,2,3}$ in $D_2, D_{2h}$), $B_1$ and $B_2$ share the same pair of Schoenflies group-type labels $G(H)$ \cite{Kopsky2006,Litvin2014} while swapping primes between secondary symmetry directions in a fixed crystal-axis setting (e.g., $B_{1u}: 6'/mmm'$ vs.\ $B_{2u}: 6'/mm'm$ in $D_{6h}$).

\FloatBarrier

\begin{figure*}[!t]
\centering
\begin{tikzpicture}[
  >=Stealth,
  node distance=0.20cm,
  stage/.style={draw, rounded corners, fill=blue!6, text width=17.4cm, inner sep=4pt, font=\footnotesize, align=flush left},
  substage/.style={draw, rounded corners, fill=blue!4, text width=8.45cm, minimum height=2.15cm, inner sep=4pt, font=\footnotesize, align=flush left},
  arrow/.style={->, thick, shorten >=1pt, shorten <=1pt}
]
\node[stage] (s1) {%
\makebox[\linewidth][s]{\textbf{Stage 1: Two-Tier Representation Encoding ($G \times \Theta_{\mathcal{T}}$ \& Sublattice $\mathcal{S}_{\mathrm{mag}}$)} \hfill \textit{Parent-Group \& Sublattice Symmetry}}\par\vspace{2pt}
$\bullet$ Sublattice Modes (Part~I): $\Gamma_{\mathrm{mag}}^{\mathrm{mat}} = \left(\Gamma_{\mathrm{ph}}\vert_{\mathcal{S}_{\mathrm{mag}}}\right) \otimes \Gamma_{\mathrm{ps}} = \bigoplus_\rho m_\rho^{\mathrm{mat}} \rho$ \hfill $\bullet$ Parent Response Space: $\Gamma_{\mathrm{ME}}^{[2]} = \Gamma_{\mathrm{pol}}^{\otimes 2} \otimes \Gamma_{\mathrm{ps}} = \bigoplus_\rho n_\rho^{\mathrm{ME}} \rho$\par\vspace{1.5pt}
$\bullet$ Two-Tier Active Intersection: $\mathcal{S}_{\mathrm{act}}^{\mathrm{mat}} = \Supp(\Gamma_{\mathrm{mag}}^{\mathrm{mat}}) \cap \Supp(\Gamma_{\mathrm{ME}}^{[2]})$ \hfill $\bullet$ Pseudoscalar Cancellation: $N(\Gamma_L) = \langle \Gamma_L, \Gamma_{\mathrm{ME}}^{[2]} \rangle = \langle \Gamma, \Gamma_{\mathrm{pol}}^{\otimes 2} \rangle$\par\vspace{1.5pt}
$\bullet$ Constitutive Channel Space: $\mathcal{M}_{\mathrm{ME}}(\Gamma_L) = \operatorname{Hom}_G(\Gamma_L, \Gamma_{\mathrm{ME}}^{[2]})$ with $N = \dim_{\mathbb{C}}\mathcal{M}_{\mathrm{ME}} = n_{\Gamma_L}^{\mathrm{ME}}$ channels for real-type $\Gamma_L \in \mathcal{S}_{\mathrm{act}}^{\mathrm{mat}}$};

\node[substage, below=0.20cm of s1.south west, anchor=north west] (s2a) {%
\textbf{Track A: Kinematic Realization (Part~I)}\par\vspace{2pt}
$\bullet$ Projection Equivalence: $\operatorname{Im} P_{\mathrm{mag}}^{\Gamma\otimes\Gamma_{\mathrm{ps}}} = \operatorname{Im} P_{\mathrm{ph}}^\Gamma$ ($P_{\mathrm{mag},mn}^{\Gamma\otimes\Gamma_{\mathrm{ps}}} = P_{\mathrm{ph},mn}^\Gamma$)\par\vspace{1.5pt}
$\bullet$ Real-Space Spin Texture: $\mathbf{M}(\mathbf{r}) = \sum_a \eta_a \boldsymbol{\phi}_a$\par\vspace{1.5pt}
$\bullet$ Nodal Invariance: Symmetry-enforced nodal manifolds};

\node[substage, below=0.20cm of s1.south east, anchor=north east] (s2b) {%
\textbf{Track B: Constitutive Projection (Part~II)}\par\vspace{2pt}
$\bullet$ Matrix-Unit Projectors: $\mathcal{P}_{k\mu}^{(\Gamma_L)}$ on field dyads $\{E_i H_j\}$\par\vspace{1.5pt}
$\bullet$ Bilinear Multiplets: $\mathbf{B}^{(n)}(\mathbf{E},\mathbf{H}) = \sum_{ij} \mathbf{C}_{ij}^{(n)} E_i H_j$\par\vspace{1.5pt}
$\bullet$ Geometric Templates: $C_{ij, k}^{(n)} = \partial^2 B_k^{(n)}/\partial E_i \partial H_j$};

\node[stage, below=0.20cm of s2a.south -| s1.center, anchor=north] (s3) {%
\makebox[\linewidth][s]{\textbf{Stage 3: Thermodynamic Invariants \& External Branch Selection} \hfill \textit{Input Magnetic Order}}\par\vspace{2pt}
$\bullet$ Symmetry-Allowed Coupling: $F_{\mathrm{me}}^{(1)} = -\sum_{n=1}^N \lambda_n I^{(n)} = -\sum_{n=1}^N \lambda_n (\mathbf{L} \cdot \mathbf{B}^{(n)}(\mathbf{E},\mathbf{H}))$\par\vspace{1.5pt}
$\bullet$ External Branch Input: Experiment or microscopic free-energy minimization selects the equilibrium order parameter $\mathbf{L}^{(0)}$};

\node[stage, below=0.20cm of s3] (s4) {%
\makebox[\linewidth][s]{\textbf{Stage 4: Macroscopic Response Emergence} \hfill \textit{Physical Response Components}}\par\vspace{2pt}
$\bullet$ Generated Tensor: $\alpha_{ij}^{(1)}(\mathbf{L}^{(0)}) = \sum_{n=1}^N \lambda_n (\mathbf{L}^{(0)} \cdot \mathbf{C}_{ij}^{(n)})$ \hfill $\bullet$ Observable Selectivity: $\Delta_{ij} \equiv \alpha_{ij} - \alpha_{ji} = 2\epsilon_{ijk} (T_\alpha)_k$\par\vspace{1.5pt}
$\bullet$ Emergent Shubnikov MPG: $M_{\mathrm{sub}}(\mathbf{L}^{(0)}) = \operatorname{Stab}_{G\times\Theta_{\mathcal{T}}}(\mathbf{L}^{(0)})$ [reducing to $H + (G-H)\mathcal{T}$ with $H = \ker\chi_L$ for 1D sign irreps]\par\vspace{1.5pt}
$\bullet$ Downstream MPG Embedding: $\operatorname{Im}(\operatorname{ev}_{\mathbf{L}^{(0)}}) \subseteq \operatorname{Fix}_{M_{\mathrm{sub}}}(V_{\mathrm{ME}})$ (1D sign irreps: isomorphism; $d_L > 1$: full or strict subspace)\par\vspace{1.5pt}
$\bullet$ Multipolar Response Components: Monopole $\theta$, Toroidal $\mathbf{T}_\alpha$, Quadrupole $\mathbf{Q}_S$};

\draw[arrow] (s1.south -| s2a.north) -- (s2a.north);
\draw[arrow] (s1.south -| s2b.north) -- (s2b.north);
\draw[arrow] (s2a.south) -- (s2a.south |- s3.north);
\draw[arrow] (s2b.south) -- (s2b.south |- s3.north);
\draw[arrow] (s3) -- (s4);
\end{tikzpicture}
\caption{Technical summary of the two-tier representation-to-response workflow connecting Part~I (Track~A) and Part~II (Track~B). The thermodynamic branch selection in Stage~3 is an external input provided by experiment or microscopic energetics; full magnetic ground-state minimization is not performed within the symmetry framework.}
\label{fig:tech_summary}
\end{figure*}
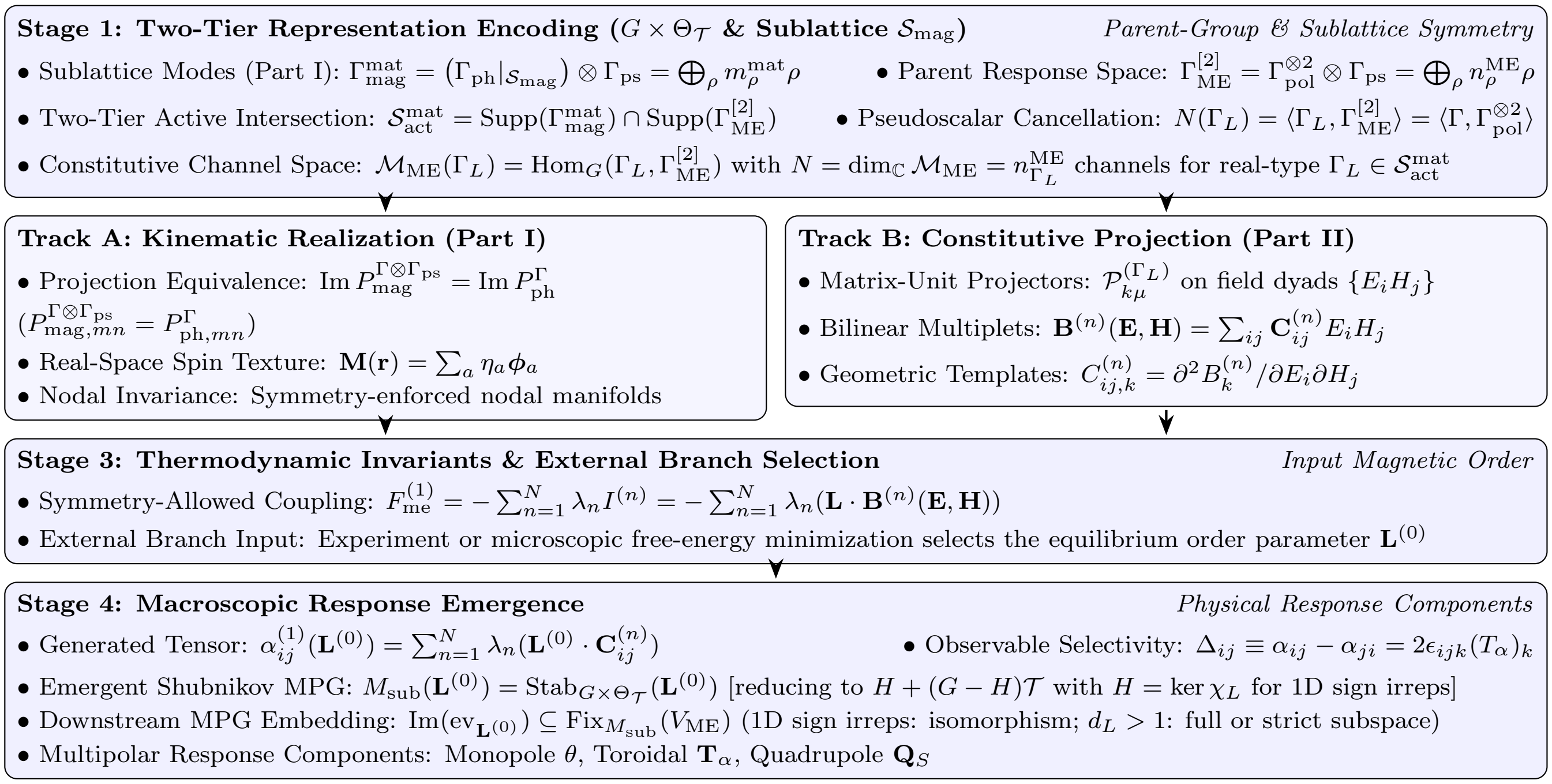

\section{Relation to Magnetic-Point-Group Tensor Constraints and Technical Summary}
\label{app:category}
\setcounter{equation}{0}
\renewcommand{\theequation}{C\arabic{equation}}
\renewcommand{\theHequation}{C\arabic{equation}}

This appendix summarizes the two linear-algebraic relations underlying the main text---the pseudoscalar correspondence between magnetic and polar coupling spaces and the exact bridge between parent-group constitutive templates and residual magnetic-point-group (MPG) tensor spaces---together with the full technical workflow diagram (Fig.~\ref{fig:tech_summary}). We use complexified representations for character-theoretic multiplicity counting and evaluate physical tensors in real Cartesian coordinates (with real channel dimensions given in Appendix~\ref{app:proof}).

\subsection{Pseudoscalar Correspondence and Two-Tier Coupling Spaces}
\label{app:category_hom}
For a parent spatial point group $G$, tensoring with the one-dimensional determinant representation $\Gamma_{\mathrm{ps}}$ ($\Gamma_{\mathrm{ps}}^{\otimes 2} \cong \mathbf{1}$) defines an involutive map $\mathcal{F}_{\mathrm{ps}}(V) = V \otimes \Gamma_{\mathrm{ps}}$ on finite-dimensional representations of $G$. Applying $\mathcal{F}_{\mathrm{ps}}$ to the second-rank polar tensor space $V_{\mathrm{pol}}^{\otimes 2}$ and to the sublattice-restricted polar displacement space $\operatorname{Im} Q_{\mathrm{mag}}$ yields the universal linear-ME response space and the material-specific sublattice magnetic space:
\begin{equation}
\begin{aligned}
V_{\mathrm{ME}} &\cong V_{\mathrm{pol}} \otimes V_{\mathrm{axial}} = \mathcal{F}_{\mathrm{ps}}\!\left( V_{\mathrm{pol}}^{\otimes 2} \right) \cong \bigoplus_{\rho \in \mathrm{Irr}(G)} \mathbb{C}^{n_\rho^{\mathrm{ME}}} \otimes V_\rho, \\
V_{\mathrm{mag}}^{\mathrm{mat}} &\equiv \Gamma_{\mathrm{mag}}^{\mathrm{mat}} = \mathcal{F}_{\mathrm{ps}}(\operatorname{Im} Q_{\mathrm{mag}}) \cong \bigoplus_{\rho \in \mathrm{Irr}(G)} \mathbb{C}^{m_\rho^{\mathrm{mat}}} \otimes V_\rho.
\end{aligned}
\label{eq:appC_VME}
\end{equation}
By Schur's lemma, the space of all $G$-equivariant linear maps from the full sublattice mode space to the response space decomposes over the two-tier active intersection $\mathcal{S}_{\mathrm{act}}^{\mathrm{mat}} = \Supp(V_{\mathrm{mag}}^{\mathrm{mat}}) \cap \Supp(V_{\mathrm{ME}})$:
\begin{equation}
\operatorname{Hom}_G\!\left( V_{\mathrm{mag}}^{\mathrm{mat}}, V_{\mathrm{ME}} \right) \cong \bigoplus_{\rho \in \mathcal{S}_{\mathrm{act}}^{\mathrm{mat}}} \operatorname{Hom}_{\mathbb{C}}\!\left( \mathbb{C}^{m_\rho^{\mathrm{mat}}}, \mathbb{C}^{n_\rho^{\mathrm{ME}}} \right).
\label{eq:appC_total_Hom}
\end{equation}
Once a single irreducible normal mode $\Gamma_L \in \mathcal{S}_{\mathrm{act}}^{\mathrm{mat}}$ is selected, its admissible leading-order constitutive maps form $\mathcal{M}_{\mathrm{ME}}(\Gamma_L) \equiv \operatorname{Hom}_G(\Gamma_L, V_{\mathrm{ME}}) \cong (V_{\mathrm{ME}} \otimes \Gamma_L^*)^G$, with complex dimension $n_{\Gamma_L}^{\mathrm{ME}}$ (matching $N$ for real-type irreps, and yielding $2n_\sigma^{\mathrm{ME}}$ real parameters for complex-type doublets per Appendix~\ref{app:proof}). Because $\mathcal{F}_{\mathrm{ps}}^2$ is the identity, whenever $\Gamma_L = \Gamma \otimes \Gamma_{\mathrm{ps}}$ corresponds to a polar representation $\Gamma$, the equivariant mapping spaces are canonically isomorphic:
\begin{equation}
\boxed{
\operatorname{Hom}_G\!\left( \Gamma \otimes \Gamma_{\mathrm{ps}}, \, V_{\mathrm{pol}}^{\otimes 2} \otimes \Gamma_{\mathrm{ps}} \right) \cong \operatorname{Hom}_G\!\left( \Gamma, V_{\mathrm{pol}}^{\otimes 2} \right)
},
\label{eq:appC_bridge_iso}
\end{equation}
establishing the untwisted channel-count identity $N(\Gamma_L) = \langle \Gamma, \Gamma_{\mathrm{pol}}^{\otimes 2} \rangle$.

\subsection{Embedding into Residual MPG Tensor Spaces}
\label{app:category_adjunction}
For a non-gray ordered magnetic descendant $M \le \mathcal{G}_{\mathrm{P}}$ with spatial projection $K = \pi_{\mathrm{spatial}}(M) \le G$, Frobenius reciprocity expresses the downstream MPG-invariant tensor subspace as:
\begin{equation}
\begin{aligned}
\operatorname{Fix}_M(V_{\mathrm{ME}}) &\cong \operatorname{Hom}_K\left(\chi_M, \operatorname{Res}_K^G V_{\mathrm{ME}}\right) \\
&\cong \operatorname{Hom}_G\left(\operatorname{Ind}_K^G \chi_M, V_{\mathrm{ME}}\right),
\end{aligned}
\label{eq:appC_Hom_Fix}
\end{equation}
where $\chi_M = \mathbf{1}_K$ for Type~I ($M=K$) and $\chi_M: K \to \{\pm 1\}$ is the alternating character with kernel $H$ for Type~III ($M = H + \mathcal{T}(K-H)$). When $G$ is ME-complete, $K$ inherits completeness (Appendix~\ref{app:proof}), ensuring $\operatorname{Fix}_M(V_{\mathrm{ME}}) \neq \{0\}$ for every non-gray descendant (Corollary~\ref{cor:nongrey}).

Equation~\eqref{eq:appC_Hom_Fix} provides the \mbox{exact} upstream-to-downstream bridge between parent-group templates and residual MPG constraints: for any one-dimensional real sign irrep $\Gamma_L = \chi_L$ of $G$ ($K=G$), evaluation on the 1D order-parameter basis yields an isomorphism $\operatorname{Hom}_G(\chi_L, V_{\mathrm{ME}}) \cong \operatorname{Fix}_M(V_{\mathrm{ME}})$. For a multidimensional irrep $\Gamma_L$ condensed along a nonzero OPD $\mathbf{L}^{(0)} \in \Gamma_L$ with isotropy subgroup $M_{\mathrm{sub}}(\mathbf{L}^{(0)})$ ($K < G$), the evaluation map $\operatorname{ev}_{\mathbf{L}^{(0)}}: \operatorname{Hom}_G(\Gamma_L, V_{\mathrm{ME}}) \hookrightarrow \operatorname{Fix}_{M_{\mathrm{sub}}(\mathbf{L}^{(0)})}(V_{\mathrm{ME}})$, $\Phi \mapsto \Phi(\mathbf{L}^{(0)})$, is injective (by irreducibility of $\Gamma_L$) but not generally surjective, because $\operatorname{Ind}_K^G \chi_{M_{\mathrm{sub}}}$ in Eq.~\eqref{eq:appC_Hom_Fix} can contain additional copies of $\Gamma_L$ and/or other parent irreps that occur in $V_{\mathrm{ME}}$; thus $\operatorname{Im}(\operatorname{ev}_{\mathbf{L}^{(0)}})$ isolates the exact leading-order constitutive subspace generated by $\Gamma_L$.

\FloatBarrier

\end{document}